# Observation of Coherent Ferron Emission and Propagation

Jeongheon Choe[1,*], Taketo Handa[1,*,†], Chun-Ying Huang[1], André Koch Liston[1], Jordan Cox[1], Jonathan Stensberg[1,2], Yongseok Hong[1], Daniel G. Chica[1], Ding Xu[1], Fuyang Tay[2], Samra Husremovic[1], Vinicius da Silveira Lanza Avelar[1], Eric A. Arsenault[1], Zhuquan Zhang[1], James McIver[2], Dmitri N. Basov[2], Milan Delor[1,†], Xavier Roy[1,†], X.-Y. Zhu[1,†]

[1] Department of Chemistry, Columbia University, New York, NY 10027, USA

[2] Department of Physics, Columbia University, New York, NY 10027, USA

**ABSTRACT. Ordered phases give rise to collective modes and quasiparticles, such as spin waves and magnons emerging from magnetic order. Extending this paradigm to ferroelectrics suggests the existence of polarization waves and their fundamental quanta, ferrons. A coherent ferron, that is a polarization wave, modulates the magnitude of electric polarization and is, thus, an amplitude (Higgs) mode of the ferroelectric order. Here we observe coherent ferrons from the pulsed laser excitation of van der Waals (vdW) ferroelectrics, $NbOI_2$ and $WO_2Br_2$. We show two complementary manifestations of the coherent ferron, the intense narrowband terahertz (THz) emission at the ferroelectric transverse optical (TO) phonon frequency and the uniaxial propagation along the polar axis as hyperbolic phonon polaritons with exceptionally long coherence times. These long-lived, uniaxial, and dipole-carrying polarization waves may find applications in narrow-band THz emission, ferronic information processing, and coherent electric control.**

In magnetic materials, the collective oscillation of magnetic axial vectors gives rise to spin waves or coherent magnons[1,2]. Similarly, the collective oscillation of polar vectors in ferroelectrics should yield electric polarization waves or coherent ferrons. While the existence of ferroelectric collective modes has long been predicted theoretically[3], particularly by Bauer and coworkers[4–8], experimental observation of coherent ferrons has remained elusive. A coherent ferron is specific to the ferroelectric TO phonon mode. It is not clear how such a polarization wave is positioned within the phonon-polariton family, particularly the highly confined hyperbolic phonon polaritons [9,10]. We address these issues using the vdW ferroelectric material $NbOI_2$.

[*] These authors contributed equally.
[†] To whom correspondence should be addressed. TH: th2922@columbia.edu; MD: md3864@columbia.edu; XR: xr2114@columbia.edu; XYZ: xyzhu@columbia.edu.

$NbOI_2$ adopts a monoclinic crystal structure in the *C*2 space group, with the polar *b*-axis and the non-polar *c*-axis in the two-dimensional (2D) plane, and the *a*-axis along the out-of-plane, stacking direction (Fig. 1a) [11,12]. Along the *b*-axis, the Nb atoms are displaced from the octahedral center positions, resulting in spontaneous polarization. Each 2D layer features robust in-plane displacive ferroelectricity at room temperature and the polarizations of consecutive layers align parallel to one another[11,12]. We show the real (Fig. 1b) and imaginary (Fig. 1c) parts of the dielectric functions along the *b*-axis (red) and the *c*-axis (blue) from THz time-domain spectroscopy (THz-TDS). This spectral region is dominated by the ferroelectric TO phonon at 3.13 THz along the *b*-axis and a weak TO mode at 3.05 THz along the *c*-axis [13].

We launch coherent ferrons in $NbOI_2$ with an ultrafast laser pulse in either an impulsive Raman or a displacive mechanism[14–17] depending on whether pump photon energy $h\nu_1$ is below or above the bandgap, respectively (Extended Data Fig. 1 for dielectric functions). A part of coherent ferron with momentum vector within the light-cone emits intense THz radiation detected in the far field. The part of coherent ferron beyond the light-cone propagates in-plane and is detected in spatially resolved measurements. We examine two structurally related vdW crystals: the ferroelectric $WO_2Br_2$ and the non-ferroelectric $TaOBr_2$ (Fig. 1a, Extended Data Fig. 2b,c). We carry out theoretical analysis and simulation to establish the connection of the propagating coherent ferron to a canalized hyperbolic phonon polariton specific to the TO mode. We use the terms "polarization wave" and "coherent ferron" interchangeably here.

**THz emission from coherent ferrons**

The first observation of the coherent ferron is THz emission at the ferroelectric TO frequency. Using $NbOI_2$ crystals of thickness $d < 1$ μm, we launch coherent ferrons by femtosecond laser pulses with polarization along the polar *b*-axis (Methods). Here we present results at temperature T = 295 K; results at cryogenic temperatures are similar (Extended Data Fig. 3a,b). We observe intense THz emission with either below-gap ($h\nu_1 = 1.55$ eV), Fig. 1d, or above-gap ($h\nu_1 = 3.1$ eV, Extended Data Fig. 3c,d) excitation. The former rules out the role of carriers or photocurrent. The initial few-cycle coherent burst at $t \leq 2$ ps is dominated by optical rectification, which yields a broadband THz emission spectrum (shaded region in Fig. 1e)[18]. Due to an extraordinarily large second-order susceptibility $\chi^{(2)}$, the efficiency of optical rectification in $NbOI_2$ is more than one order of magnitude larger than that in the standard THz emitter, ZnTe [18].

This initial burst in broadband THz is followed by persistent coherent oscillation, which gives an exceptionally sharp peak at 3.13 ± 0.01 THz from Fourier transform with a full width at half maximum (FWHM) of 0.04 THz (Fig. 1e), limited by the measurement time window. The spectral amplitude at 3.13 THz is larger than that of the broadband optical rectification, highlighting the exceptional efficiency. The 3.13 THz mode corresponds to the ferroelectric TO phonon, as shown by the dielectric function in Fig. 1b,c. Density functional theory (DFT) calculations show that this TO mode modulates the length of O-Nb-O bonds[13] (Fig. 1e, inset) and, thus, spontaneous polarization. The magnitude of the 3.13 THz field scales linearly with excitation intensity (Extended Data Fig. 4), in agreement with an impulsive Raman mechanism. The coherent ferron at the TO frequency also modulates second harmonic generation (SHG) (Extended Data Fig. 5), whose intensity reflects the order parameter—that is, the amplitude of macroscopic polarization which breaks inversion symmetry.

We perform experiments on two vdW materials structurally similar to $NbOI_2$: the ferroelectric $WO_2Br_2$ and the non-ferroelectric $TaOBr_2$ (Fig. 1a). Single crystals of $WO_2Br_2$ and $TaOBr_2$ are synthesized through chemical vapor transport detailed in Supplementary Text 1 (see Extended Data Figs. 2a-c for photos of as-grown single crystals of $NbOI_2$, $WO_2Br_2$ and $TaOBr_2$). $WO_2Br_2$ crystallizes in the non-centrosymmetric space group *Imm2*, exhibiting unequal W–O bond lengths along the *c*-axis (Supplementary Tables 1,2), giving rise to spontaneous polarization. The polarizations of consecutive vdW layers align in parallel, similar to $NbOI_2$. In contrast, $TaOBr_2$ adopts the centrosymmetric *Immm* space group (Supplementary Tables 3,4); consequently, $\chi^{(2)}$ vanishes and prohibits impulsive Raman excitation of the TO mode. The THz emission of $WO_2Br_2$ shows persistent coherent oscillation, similar to that in $NbOI_2$. Fourier transform of the long-lived ($t$ > 2 ps) emission gives narrowband THz emission at 2.76 ± 0.01 THz (inset in Fig. 1f). As in $NbOI_2$, we attribute the narrowband emission from $WO_2Br_2$ to the ferroelectric TO mode. The amplitude of THz emission from $WO_2Br_2$ is generally lower than that from $NbOI_2$, in agreement with the ~5X lower $\chi^{(2)}$ in the former, as estimated from SHG. The polarity of the THz signal from $NbOI_2$ flips with the direction of ferroelectric polarization for both the initial burst and the long-lived coherent oscillation, Fig. 1g, further supporting the $\chi^{(2)}$ origin of THz emission.

**Extremely long-lived, hypersonic, and uniaxial propagation**

The second property of the coherent ferron is its uniaxial propagation with group velocities far exceeding that expected from the TO phonon dispersion. We excite with a focused laser pulse (Gaussian width ~ 1.6 μm) and determine coherent ferron propagation using a time-delayed ($\Delta t$) and spatially-separated ($\Delta x$) probe pulse ($h\nu_2$ = 1.8−2.5 eV) in transient reflectance[13], $\Delta R/R_0$, where $R_0$ is reflectance without pump (Methods). Figure 2 shows the propagation at $T$ = 3.8 K; similar results are obtained at T = 295 K (Extended Data Fig. 6). We use below gap ($h\nu_1$ = 1.77 eV; Fig. 2) or above gap ($h\nu_1$ = 2.4 eV; Extended Data Fig. 7) excitation and observe similar propagations. Fig. 2a,b present data for $NbOI_2$ flake thickness $d$ = 224 nm and 96 nm; results at other thicknesses are in Extended Data Fig. 8. The $\Delta R/R_0$ signal is plotted as a function of $\Delta t$ at $\Delta x$ = 0.0, 2.0, 4.0, and 6.0 μm along the polar $b$-axis. Fourier transforms reveal a dominant peak at 3.132 ± 0.005 THz, Fig. 2c.

To determine the propagation velocities, we apply short-time Fourier transform (STFT) to the time traces in Fig. 2a,b. While the frequency of 3.132 ± 0.005 THz remains largely constant, the oscillation amplitudes peak at longer Δt with increasing Δx (Fig. 2d). We obtain the wavepacket arrival time ($\Delta t_{WP}$) as the STFT amplitude reaches maximum at each $\Delta x$, marked by colored arrows in Fig. 2a,b. We determine the group velocity ($v_g$) from a linear fit of $\Delta t_{WP}$ versus $\Delta x$. The $v_g$ values (50 - 120 km·s$^{-1}$), Fig. 2e, scale positively with thickness, in agreement with the solid line from theoretical analysis below. The group velocities are over 2-4 orders of magnitude larger than those of phonons (Supplementary Table 5). In control experiments, we find no measurable signal when the probe is displaced from the pump spot along the non-polar $c$-axis.

The uniaxial polarization waves exhibit exceptionally long coherence lifetimes. At small Δx, the apparent decays of oscillatory signal may come from population loss due to wavepacket fractions leaving the probe spot and, for the slowest moving coherent ferrons at $q \leq 0.066$ μm$^{-1}$, to THz emission. We estimate coherence times from time domain fits at $\Delta x$ = 6 μm. Propagation simulations (Extended Data Fig. 8i) show that the extracted coherence times are lower bounds. From time domain fits, as illustrated in Extended Data figure 8g,h, we obtain mean coherence times of $\tau_{coh}$ ~ 100-300 ps at 4 K. At T = 295 K, the fits give mean coherence times of $\tau_{coh}$ ~ 12-30 ps [13]. The range reflects sample-to-sample variation, particularly the quality of crystal surfaces.

As established below, these uniaxial polarization waves are canalized hyperbolic phonon-polaritons. In the seminal paper on $\alpha$-$MoO_3$, the directional and in-plane hyperbolic phonon-polariton in the lower Reststrahlen band has a lifetime $\tau = 1.9 \pm 0.3$ ps [19], increased to $\tau = 3.1 \pm 0.4$ ps with monochromatic excitation[20]. Canalized phonon-polaritons have been observed in $LiV_2O_5$ and $\alpha$-$MoO_3$ with $\tau \sim 2$ ps [21,22]. In $YVO_4$, the longest lifetime of directional hyperbolic phonon polariton is $\tau = 5.55$ ps at $T = 150$ K[23]. The coherence times observed for coherent ferrons in $NbOI_2$ are an order of magnitude longer than those previously reported. The dielectric functions in Figs. 1b and 1c feature an extremely sharp TO resonance, with FWHM = 8±2 GHz, which is an order of magnitude smaller than those of typical polar semiconductors, suggesting much suppressed anharmonic coupling to other phonon modes in $NbOI_2$.

**Direct spatiotemporal imaging of the polarization waves**

We directly image the propagation of coherent ferrons in $NbOI_2$ using stroboscopic scattering microscopy (stroboSCAT, Fig. 3a), which tracks quasiparticle propagation with femtosecond and sub-50 nm spatiotemporal precision by detecting small excitation-induced changes to the refractive index[24]. Following focused pump excitation (~730 nm diameter, ~160 fs) at 2.4 eV, a ~30 fs widefield probe at 2.33 eV is imaged in a backscattering geometry on an array camera to capture coherent ferron propagation as function of $\Delta$t. Fig. 3b displays the stroboSCAT images from a 240 nm thick $NbOI_2$ flake after application of a Fourier bandpass filter centered at 3.1325 ± 0.0025 THz (see Extended Data Fig. 9 for different bandpass frequencies allowing us to resolve different sections of the ferron dispersion). Fourier bandpassing also removes the exponentially-decaying excitonic population background at the excitation spot arising from above-gap excitation. The signal phase rapidly oscillates at a frequency of 3.132 ± 0.005 THz (see Fig. 3c at 50 fs intervals). Images of STFT amplitudes in Fig. 3b are shown in Fig. 3d, clearly displaying uniaxial propagation along the polar *b*-axis. In addition to the fast-moving wavefront, we also observe persistent and slower moving features near the excitation spot. This slower component likely originates from sub-bands with lower $v_g$, as supported by simulations presented below.

The striking spatiotemporal evolution of polarization waves is best visualized in Supplementary Video 1. Extended Data Fig. 9a,b show stroboSCAT images at $\Delta t$ = 40 ps for Fourier band-passes at adjacent frequencies. Beyond the peak TO frequency, the contrast intensity is diminished and the characteristic profile is lost at frequencies approaching the LO phonon (3.220

THz). The spatial profiles agree well with the simulations shown below (and Extended Data Fig. 9 for different frequencies), including the anisotropic and ultrafast propagation of the wavefront and the frequency dependent propagation profiles. Fig. 3e plots the time at which the STFT amplitude peaks at positions away from the excitation spot. We extract a propagation velocity of 110.6 ± 1.0 km/s for a 240 nm thick flake, in agreement with data shown in Fig. 2e.

**Simulations: the connection of coherent ferrons to phonon-polaritons**

In the ferron framework, the ferron quasiparticle emerges from the TO phonon in the presence of both anharmonicity and broken inversion symmetry[4–8,25–27]. A ferron carries electric polarization, $\delta\boldsymbol{p}$, which suppresses the static polarization $\boldsymbol{P}$. At the lowest order, the Landau free energy for polarization along the x-axis is:

$$F = \frac{\alpha}{2}\boldsymbol{P}_x^2 + \frac{\beta}{4}\boldsymbol{P}_x^4 - \boldsymbol{E}\cdot\boldsymbol{P} \quad (1),$$

where the Landau parameters ($\alpha < 0$ and $\beta > 0$) describe the familiar double-well potential, while $\boldsymbol{E}$ is electric field. At equilibrium with $\boldsymbol{E} = 0$, we have $\frac{\partial F}{\partial \boldsymbol{P}} = 0$ and spontaneous polarization $|\boldsymbol{P}_0| = \sqrt{-\alpha/\beta}$. The equation of motion of polarization fluctuation along the longitudinal $\boldsymbol{P}_0$ direction can be described by the Landau-Khalatnikov-Tani (LKT) equation [28]:

$$m_p \frac{\partial^2 \boldsymbol{P}}{\partial t^2} + \gamma \frac{\partial \boldsymbol{P}}{\partial \boldsymbol{t}} = -\frac{\partial F}{\partial \boldsymbol{P}} + E(t) \quad (2),$$

where $m_p = 1/(\varepsilon_0 \omega_p^2)$ is the effective mass, $\omega_p$ is the angular frequency of ionic motion, $\gamma$ is a phenomenological damping constant, and $E(t)$ is the dynamic electric field. With negligible dissipation ($\gamma \rightarrow 0$), the elementary electric dipole of a ferron becomes [5]:

$$\delta\boldsymbol{p}_{\boldsymbol{q}} = \frac{\hbar}{m_p \omega_{\boldsymbol{q}}} \frac{3\beta}{2\alpha} \cdot \boldsymbol{P}_{\boldsymbol{0}} \quad (3),$$

where $\omega_{\boldsymbol{q}}$ and $\boldsymbol{q}$ are the ferron frequency and wave vector, respectively. A coherent ferron is a wave packet $\delta p_{\boldsymbol{q}}(\mathbf{r}, \mathrm{t}) = \sum_{\boldsymbol{q}'} \delta p_{\boldsymbol{q}'}$ summed in a finite momentum window around $\boldsymbol{q}$. Assuming a plane wave form,

$$\delta p_{\boldsymbol{q}}(\mathbf{r}, \mathrm{t}) \propto P_0 \cdot e^{-i(\omega_q \cdot t - \boldsymbol{q}\cdot\boldsymbol{r})} \quad (4),$$

we obtain the coherent ferron contribution to THz emission, $E_{THz} \propto \frac{\partial^2 P}{\partial t^2} = \left[\frac{\partial^2(\delta p_q)}{\partial t^2}\right]$, oscillating at $\omega_{\boldsymbol{q}}$ with an amplitude proportional to $P_0$. At the long wavelength limit where the coherent ferron momentum matches the far field photon momentum, $\omega_{\boldsymbol{q}}$ approaches $\omega_{TO}$ at the $\Gamma$ point.

While THz emission at the longitudinal optical (LO) phonon frequency is well established for optically pumped non-centrosymmetric semiconductors [29–31], TO emission has not been observed. Instead, TO phonons typically appear as absorption dips rather than emission peaks. The large dielectric function at the TO frequency can lead to re-absorption, that is, destructive interference, as the field propagates through a crystal. To confirm this, we measure a thin ZnTe crystal, where the TO phonon appears as a dip in the broad THz emission spectrum generated from optical rectification (Extended Data Fig. 10). We hypothesize that the intense THz emissions from the ferroelectric TO modes in $NbOI_2$ and $WO_2Br_2$ arise from large nonlinear contributions to the time-dependent polarization at $\omega_{TO}$ due to coherent ferrons (equation 4).

Next, we address the uniaxial propagation from analyzing the dynamics of polarization fluctuation mediated by long-range dipolar coupling, as described by Maxwell's equations[4–8,25–27]. In analogous layered magnets [32–35], the long-range magnetic dipolar interaction is found to be responsible for a three order of magnitude increase in coherent magnon velocity in the vdW antiferromagnet CrSBr [36]. As detailed in the Methods, from equations (2) and (4), we obtain the dielectric function:

$$\varepsilon_{xx} = 1 + \left(-\frac{\omega^2}{\omega_p^2} + K_x - i\varepsilon_0\gamma\omega\right)^{-1} \quad (5),$$

where $K_x = -2\varepsilon_0\alpha$ is the stiffness of polarization fluctuation determined by the Landau parameter ($\alpha < 0$) [7,8]. Fitting to the experimental data along the polar axis in Figs. 1b and 1c gives $\omega_p/2\pi$ = 3.132 THz. Within the electrostatic limit, $\nabla \times \boldsymbol{E} = 0,$ and with $\nabla \cdot \boldsymbol{D} = 0$ ($\boldsymbol{D}$ is displacement field), we have

$$\varepsilon_{xx}{q_x}^2 + \varepsilon_{yy}{q_y}^2 + \varepsilon_{zz}{q_z}^2 = 0 \quad (6)$$

and obtain two branches of dispersions (see Methods):

$$\omega_{n,\pm}(\boldsymbol{q}) = \frac{\omega_p}{\sqrt{2}}\sqrt{(1+K_x) \pm \sqrt{(1-K_x)^2 + \frac{4K_x q_x^2}{q_x^2+q_y^2+q_{z,n}^2}}} \quad (7)$$

where $q_{z,n}$ represents quantization ($n$ = 0, 1, 2…) along the surface normal due to finite sample thickness. The (+) branch is in the THz region probed here. The frequency of the (−) branch is two orders of magnitude lower and difficult to detect. Analogous to coherent magnons, coherent ferrons exhibit both bulk and surface modes[7,8]. Here we focus on the bulk mode whose group velocity scales with sample thickness, as seen in experiments.

Fig. 4a plots the dispersion of $\omega_{0,+}(\boldsymbol{q})$ along in-plane $q_x$ and $q_y$. The dispersion is strongly dispersive along the polar $q_x$ (at $q_y = 0$) and non-dispersive along the non-polar $q_y$ (at $q_x = 0$). The darkest blue curve ($\omega_0$) in Fig. 4b is a cut at $q_y = 0$, showing the dispersion from TO to LO. The dispersion is dominated by that of the ferron only at the small $q_x$ limit (< 5 $\mu m^{-1}$). It approaches the much flatter LO dispersion at $q_x$ >> 5 $\mu m^{-1}$. As we show in Method (Extended Data equation 19), for a slab with confinement in the surface normal direction (z), there are a group of sub-bands ($\omega_0$, $\omega_1$…) with quantized $q_{z,n}$ values. With increasing $q_{z,n}$, the ferron dispersion along $q_x$ flattens and $v_g$ decreases. As the sample gets thinner, $q_z$ increases, flattening the ferron dispersion and reducing $v_g$, giving rise to an approximate linear relationship between $v_g$ and $d$ (Methods), shown in Fig. 2e. This confinement effect is analogous to that of magnetostatic spin waves in magnetic thin films[34–36].

The ferron formalism is fundamentally related to the macroscopic dielectric function. A unique feature of a coherent ferron is that it transports electric dipoles, while a phonon polariton in non-ferroelectrics does not. In $NbOI_2$, the real part of the dielectric function between TO and LO is negative, $\mathrm{Re}(\varepsilon_{xx}) < 0$, while that in the nonpolar direction is positive $\mathrm{Re}(\varepsilon_{yy}) > 0$ (Fig. 1b). $\mathrm{Re}(\varepsilon_{zz})$ is not measured here, but the absence of phonon modes normal to the surface in this frequency range [13] suggests $\mathrm{Re}(\varepsilon_{zz}) > 0$. The resulting dielectric tensor is that of an in-plane hyperbolic phonon polariton [9,10]. Indeed, equation (6) for a coherent ferron at the electrostatic limit is the same as that of a phonon-polariton under high momentum approximation [9,10]. More specifically, the coherent ferron is a canalized phonon polariton [21,22] at $\omega_{TO}$, the lower boundary of the hyperbolic window. As detailed in Method, we carry out hyperbolic phonon polariton simulation and obtain the imaginary part of the reflection coefficient for $p$ polarization, $\mathrm{Im}(r_p)$. The resulting dispersions for multiple phonon-polariton modes due to quantized $q_{z,n}$ are shown in Fig. 4c, in excellent agreement with those of analytical approximations of coherent ferrons in Fig. 4b.

When a ferron wavepacket is excited by a focused laser spot, the excitation amplitude of in-plane wavevectors can be represented by a Gaussian with width of the order of $\mu m^{-1}$ (green curve in Fig. 4b). As shown in Fig 4d,e,f, we obtain the spatiotemporal map of propagating ferron amplitude from

$$\delta P(x,y,t) = A_0 \sum_n \int d^2\boldsymbol{q} e^{-\frac{q^2\sigma^2}{2}} e^{-i\omega_n(\boldsymbol{q})t+i\boldsymbol{q}\cdot\boldsymbol{r}} \quad (8),$$

where $A_0$ is the initial excitation amplitude with laser spot size $\sigma$ at $\Delta t = 0$ ps, with parameters $K_x = 0.063$, $q_{z,n} = 2 + 13n\ \mu m^{-1} (n = 0, 1)$ , $\sigma = 0.7\ \mu m$ (see Methods). From the simulated spatiotemporal maps, we calculate $v_g \sim 160$ km/s along the polar axis by tracking the maximum wave amplitude. The simulated thickness-dependent group velocities (solid line in Fig. 2e) are in excellent agreement with experimental values in Fig. 2e and Fig. 3e.

We point out that propagation inside $NbOI_2$ and emission into free space are complementary aspects of the coherent ferron. The free space photon wavevector at 3.13 THz is $2\pi/\lambda = 0.066\ \mu m^{-1}$, which is near the center (~0.3%) of the $q$ range in Fig. 4. Thus, a small portion of the coherent ferron wavepacket at the low $q$ limit emits THz into free space, while the majority of the wavepacket at $q > 0.066\ \mu m^{-1}$ is confined within the crystal slab and propagates as hyperbolic phonon polaritons. The frequency of the propagating portion should be higher than that of the emitting portion, and this difference is resolved in the propagation experiments as a slight blueshift of ~ 10 GHz for increasing pump-probe separation, as shown in Extended Data Figs. 6b, 8b, 8f.

**Summary**

The polarization waves in $NbOI_2$ exhibit superior properties. They transport electric dipoles at the THz TO frequency and at hypersonic velocities, are canalized along the polar axis, and possess long coherence times of 20 ± 7 ps and 220 ± 80 ps at room temperature and 4 K, respectively. A uniaxial polarization wave at the ferroelectric TO frequency modulates the ferroelectric order and is, thus, a coherent ferron [4–8,25–27]. Hyperbolic phonon polaritons discovered so far in other vdW materials[9,10] do not transport electric dipoles and possess much shorter lifetimes limited by their dielectric functions. Phonon-polaritons, with frequencies away from the TO mode and group velocities close to the speed of light, have been reported in three-dimensional oxide perovskite ferroelectrics [37,38]. They also carry electric dipoles, but their transport is not hyperbolic and not confined uniaxially. The uniaxial, long-lived, and dipole carrying polarization waves in $NbOI_2$ may be alternatives to spin waves[2,39–41] for novel applications in information processing and control. Moreover, the 2D vdW nature of $NbOI_2$ and $WO_2Br_2$ may greatly expand the applications of these ferroelectric materials when they are integrated into vdW heterostructures.

**ACKNOWLEDGEMENTS.** Research on coherent ferrons is supported by the US Army Research Office under grant number W911NF-23-1-0056. Material synthesis and stroboSCAT imaging were supported in part by the Materials Science and Engineering Research Center (MRSEC) on Precision Assembly of Quantum Materials (PAQM) through NSF award DMR-2011738. Structural characterization and additional simulations and theoretical analysis was supported by the Air Force Office of Scientific Research under award number FA9550-22-1-0389. Development of cryogenic pump-probe spectro-microscopy methodologies was supported in part by DOE-BES under award DE-SC0024343. Instrument development for stroboSCAT and dielectric function measurements were supported by the NSF under grant numbers DMR-2115625 and CHE-2203844, respectively (M.D.). This research utilized instrumentation provided by the Programmable Quantum Materials, an Energy Frontier Research Center funded by the U.S. Department of Energy (DOE), Office of Science, Basic Energy Sciences (BES), under award DE-SC0019443. EAA and SH acknowledge support from the Simons Foundation as Junior Fellows in the Simons Society of Fellows (965526 and SFI-MPS-SFJ-00006092). CYH acknowledges support from the Taiwan-Columbia scholarship funded by the Ministry of Education of Taiwan and Columbia University. We thank David R. Reichman and Zhi-Hao Cui for helpful discussions and Chiara Trovatello for help with SHG measurement.

**AUTHOR CONTRIBUTIONS**. X.Y.Z., X.R., T.H., and J.C. conceived this work. T.H. and C.Y.H. carried out THz emission measurements. J.C. carried out the pump-probe propagation experiments with help from C.Y.H. and E.A.A. J.C. carried out coherent ferron simulations. A.K.L. and Y.H. carried out stroboSCAT imaging experiments under the supervision of M.D. J.C. and D.G.C. carried out crystal synthesis and X-ray scattering analysis. C.Y.H. prepared the exfoliated samples. D.X. and C.Y.H. determined the dielectric function of $NbOI_2$. F.T. carried out phonon-polariton simulation under the supervision of DNB. S.H. performed TEM measurements. VSLV prepared the ZnTe crystals. T.H., J.S. and Z.Z. developed nonlinear optical analysis of THz emission, under the supervision of X.Y.Z. and J.M. X.Y.Z. and X.R. supervised the project. The manuscript was prepared by J.C., T.H., and X.Y.Z. in consultation with all other authors. All authors read and commented on the manuscript.

## Competing Interests

The authors declare no competing interests.

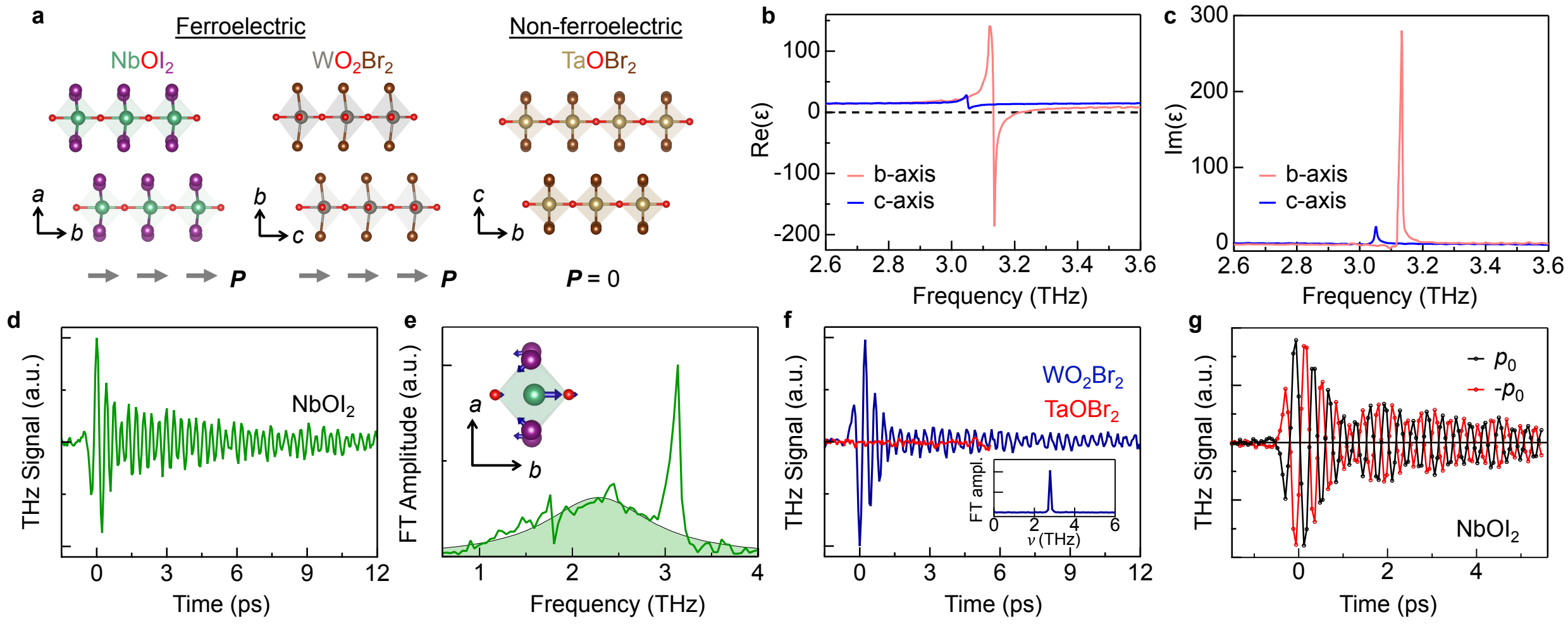

**Fig. 1. THz emission from polarization waves.** (a) Crystal structures of ferroelectric materials $NbOI_2$ and $WO_2Br_2$ and centrosymmetric material $TaOBr_2$. (b, c) Real (b) and imaginary (c) parts of the dielectric function along the in-plane polar (red) and non-polar (blue) axis in $NbOI_2$. (d) Time-domain signal of THz emission from a 650 nm thick $NbOI_2$ crystal excited by a laser pulse at $h\nu_1 = 1.55$ eV (below gap). (e) Frequency domain signal (Fourier transform) for $NbOI_2$. The inset shows the atomic motion corresponding to the 3.13 THz TO phonon mode. The green, red, and purple atoms represent niobium, oxygen, and iodine, respectively. The shaded region is the initial optical rectification. (f) Time-domain THz signal for the ferroelectric $WO_2Br_2$ (blue) and centrosymmetric $TaOBr_2$ (red) excited by a laser pulse at $h\nu_1 = 1.55$ eV. Fourier transform of the THz time domain signal after 2 ps from $WO_2Br_2$ gives a narrowband peak at 2.76 THz (inset). (g) A comparison of time-domain THz emission with ferroelectric polarization of $\mathbf{P_0}$ (black) and $\mathbf{-P_0}$ (red) achieved via 180° sample rotation. All spectra are obtained at room temperature.

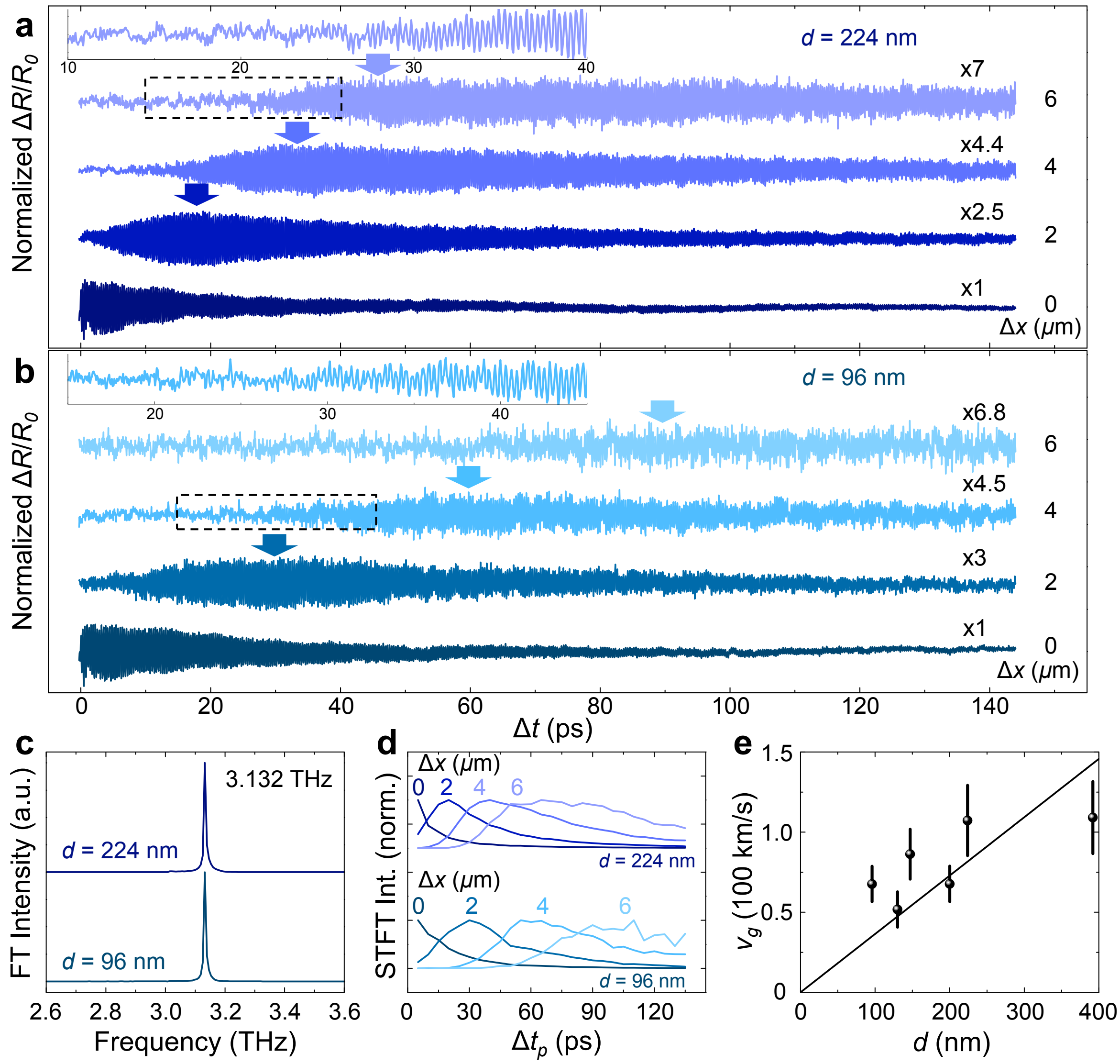

**Fig. 2. Hypersonic propagation of polarization waves.** (a,b) Transient reflectance $\Delta R/R_0$ (peak intensity normalized) as a function of pump-probe time delay (Δt) for (a) 224 nm thick and (b) 96 nm thick $NbOI_2$ at the indicated pump-probe spatial separation along the polar axis, Δx = 0, 2, 4, 6 μm, respectively. Colored arrows indicate the wavepacket arrival time. Insets are magnified plots for dashed line boxes. Incoherent background signals are subtracted. (c) Fourier transform of the data for $\Delta x = 0$ in panel (a) and (b). (d) Short-time window Fourier transform (STFT) of the transient reflectance data in (a) and (b) obtained with time window size 10 ps and step 5 ps. The normalized amplitudes of the 3.13 THz signal are shown for Δx = 0, 2, 4, 6 μm at the two sample thicknesses, d = 224 nm and 96 nm. The Δt values at peak STFT amplitudes for varying Δx are used to obtain group velocities. (e) Group velocities (symbols) of polarization waves obtained from STFT analysis for five samples with different thicknesses. Error bars are estimated from fittings to $v_g = \Delta x/\Delta t_p$. The solid line is a linear fit based on the ferron or hyperbolic phonon polariton formalism. All experiments are done at a sample temperature of 3.8 K. Pump $h\nu_1$ = 1.77 eV (fluence, 1.9 mJ/cm$^2$); probe pulse at $h\nu_2$ = 1.8–2.5 eV (fluence, 6.5 μJ/cm$^2$).

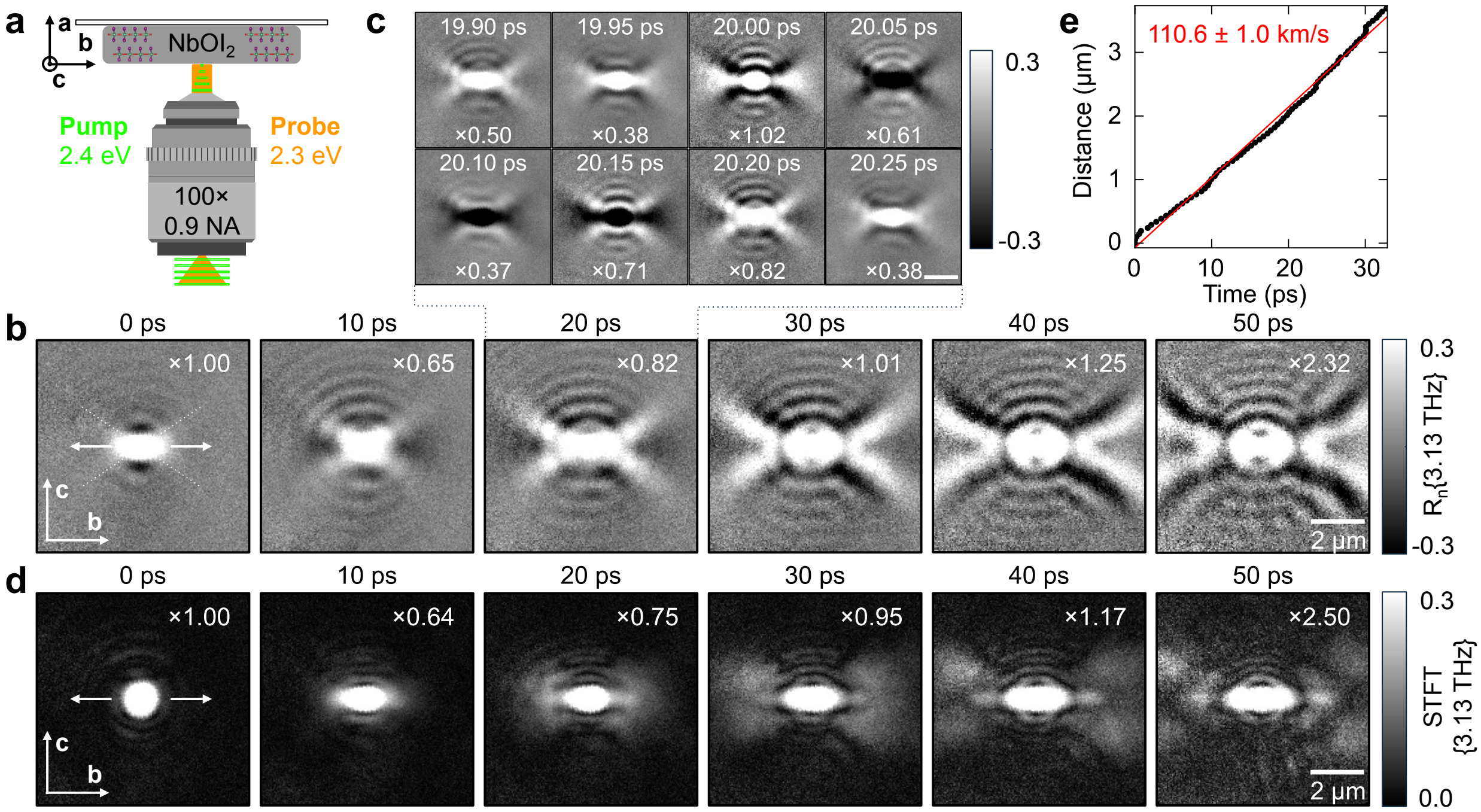

**Fig. 3. Spatiotemporal imaging of polarization wave.** (a) stroboSCAT schematic for imaging ferrons. (b) StroboSCAT images of ferron propagation in a 240 nm thick $NbOI_2$ flake at 4 K after application of a Fourier bandpass filter centered at 3.1325 ± 0.025 THz. Only images with peak positive phase are shown here (normalization factor shown on top-right for each time-delay; contrast shown on scale bar). (c) StroboSCAT images in 50 fs increments around 20 ps pump-probe delay showing the oscillatory nature of the signal. (d) STFT amplitudes from the experiment shown in panels b (normalization factor shown on top-right for each time-delay). (e) Propagation of the peak STFT amplitude away from the excitation spot (at 3.1325 ± 0.0025 THz) along polar *b*-axis. All scale bars, 2 μm.

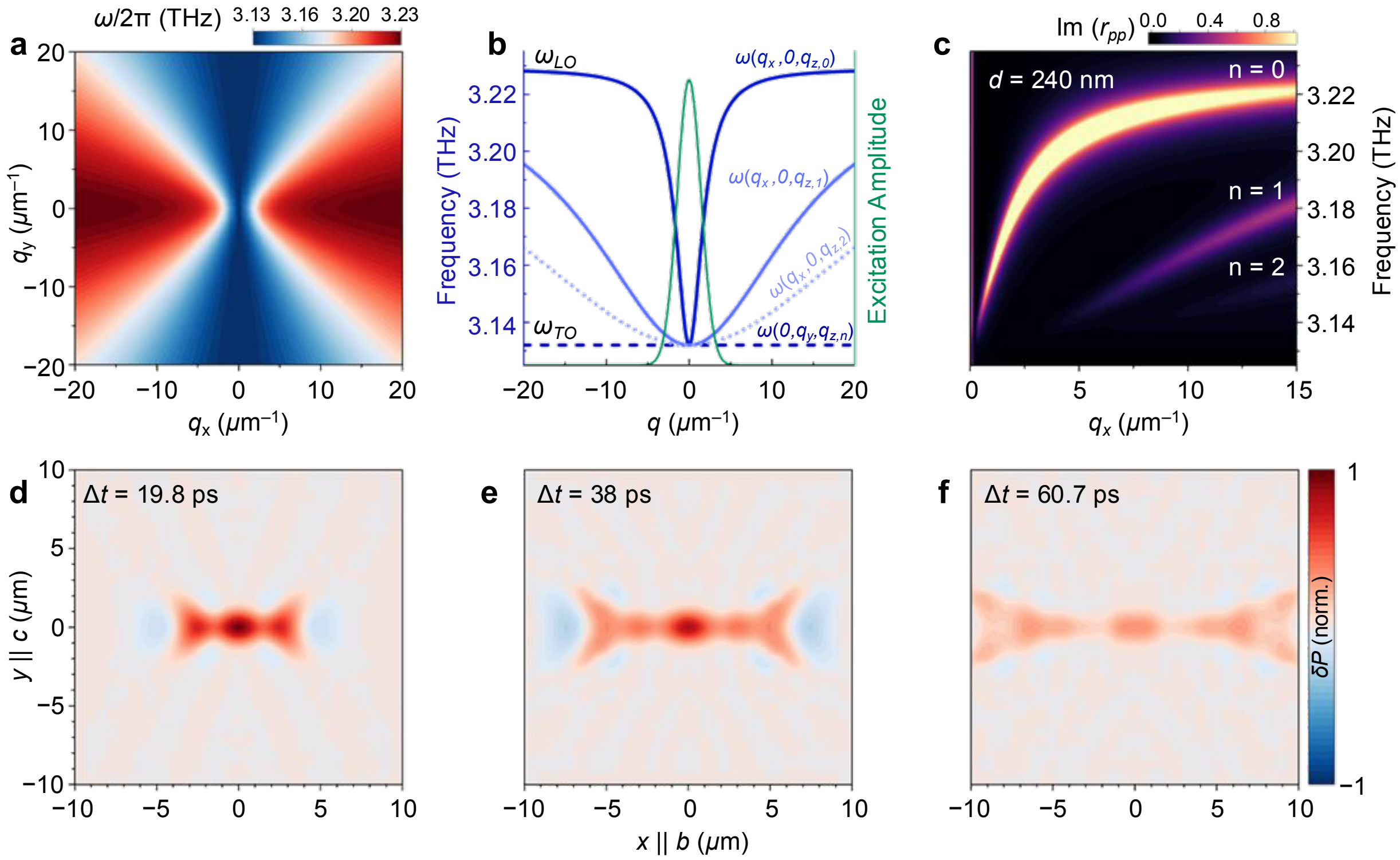

**Fig. 4. Calculated ferron dispersion and transport.** (a) Ferron dispersion at $q_{z,0} = 2\ \mu m^{-1}$. (b) Linecut at $q_y = 0$ (solid blue curve) and $q_x = 0$ (dashed blue curve) with $q_{z,n} = 2 + 13n\ \mu m^{-1}$ (n = 0, 1, 2, 3). Pump-induced excitation amplitude is shown as solid green curve. (c) Ferron dispersion obtained from hyperbolic phonon polariton simulation with multiple bands ($n$ = 0, 1, 2) from confinement in the z-direction. Sample thickness $d$ = 240 nm and substrate with 285 nm thick $SiO_2$ on Si are used. (d,e,f) Calculated spatiotemporal map of ferron transport with $n$ = 0, 1 bands at time delays $\Delta t$ = 19.8, 38, and 60.7 ps.

## References

1. Van Kranendonk, J. & Van Vleck, J. H. Spin waves. *Rev. Mod. Phys.* **30**, 1 (1958).
2. Chumak, A. V, Vasyuchka, V. I., Serga, A. A. & Hillebrands, B. Magnon spintronics. *Nat. Phys.* **11**, 453–461 (2015).
3. de Gennes, P. G. Collective motions of hydrogen bonds. *Solid State Commun.* **1**, 132–137 (1963).
4. Bauer, G. E. W., Iguchi, R. & Uchida, K. Theory of transport in ferroelectric capacitors. *Phys. Rev. Lett.* **126**, 187603 (2021).
5. Tang, P., Iguchi, R., Uchida, K. & Bauer, G. E. W. Excitations of the ferroelectric order. *Phys. Rev. B* **106**, L081105 (2022).
6. Bauer, G. E. W. *et al.* Polarization transport in ferroelectrics. *Phys. Rev. Appl.* **20**, 050501 (2023).
7. Zhou, X.-H. *et al.* Surface ferron excitations in ferroelectrics and their directional routing. *Chinese Physics Letters* **40**, 087103 (2023).
8. Rodríguez-Suárez, R. L. *et al.* Surface and volume modes of polarization waves in ferroelectric films. *Phys. Rev. B* **109**, 134307 (2024).
9. Galiffi, E. *et al.* Extreme light confinement and control in low-symmetry phonon-polaritonic crystals. *Nat. Rev. Mater.* **9**, 9–28 (2024).
10. Wang, H. *et al.* Planar hyperbolic polaritons in 2D van der Waals materials. *Nat. Commun.* **15**, 69 (2024).
11. Abdelwahab, I. *et al.* Giant second-harmonic generation in ferroelectric NbOI2. *Nat. Photonics* **16**, 644–650 (2022).
12. Wu, Y. *et al.* Data-driven discovery of high performance layered van der Waals piezoelectric NbOI2. *Nat. Commun.* **13**, 1884 (2022).
13. Huang, C.-Y. *et al.* Coupling of electronic transition to ferroelectric order in a 2D semiconductor. *Nat. Commun.* **16**, 1896 (2025).
14. Yan, Y., Gamble Jr, E. B. & Nelson, K. A. Impulsive stimulated scattering: General importance in femtosecond laser pulse interactions with matter, and spectroscopic applications. *J. Chem. Phys.* **83**, 5391–5399 (1985).
15. Cheng, T. K. *et al.* Impulsive excitation of coherent phonons observed in reflection in bismuth and antimony. *Appl. Phys. Lett* **57**, 1004 (1990).
16. Zeiger, H. J. *et al.* Theory for displacive excitation of coherent phonons. *Phys. Rev. B* **45**, 768 (1992).
17. Merlin, R. Generating coherent THz phonons with light pulses. *Solid State Commun.* **102**, 207–220 (1997).
18. Handa, T. *et al.* Terahertz emission from giant optical rectification in a van der Waals material. *Nat. Mater.* **24**, 1203–1208 (2025).
19. Ma, W. *et al.* In-plane anisotropic and ultra-low-loss polaritons in a natural van der Waals crystal. *Nature* **562**, 557–562 (2018).
20. de Oliveira, T. V. A. G. *et al.* Nanoscale-confined terahertz polaritons in a van der Waals crystal. *Advanced Materials* **33**, 2005777 (2021).
21. F. Tresguerres-Mata, A. I. *et al.* Observation of naturally canalized phonon polaritons in LiV2O5 thin layers. *Nat. Commun.* **15**, 2696 (2024).

22. Terán-García, E. *et al.* Real-Space Visualization of Canalized Ray Polaritons in a Single Van der Waals Thin Slab. *Nano Lett.* **25**, 2203–2209 (2025).
23. Liu, L. *et al.* Long-range hyperbolic polaritons on a non-hyperbolic crystal surface. *Nature* **644**, 76–82 (2025).
24. Delor, M., Weaver, H. L., Yu, Q. & Ginsberg, N. S. Imaging material functionality through three-dimensional nanoscale tracking of energy flow. *Nat. Mater.* **19**, 56–62 (2020).
25. Bauer, G. E. W., Tang, P., Iguchi, R. & Uchida, K. Magnonics vs. ferronics. *J. Magn. Magn. Mater.* **541**, 168468 (2022).
26. Tang, P., Iguchi, R., Uchida, K. & Bauer, G. E. W. Thermoelectric polarization transport in ferroelectric ballistic point contacts. *Phys. Rev. Lett.* **128**, 047601 (2022).
27. Rodríguez-Suárez, R. L. *et al.* Surface and volume modes of polarization waves in ferroelectric films. *Phys. Rev. B* **109**, 134307 (2024).
28. Sivasubramanian, S., Widom, A. & Srivastava, Y. N. Physical kinetics of ferroelectric hysteresis. *Ferroelectrics* **300**, 43–55 (2004).
29. Kuznetsov, A. V & Stanton, C. J. Coherent phonon oscillations in GaAs. *Phys. Rev. B* **51**, 7555 (1995).
30. Dekorsy, T. *et al.* Emission of submillimeter electromagnetic waves by coherent phonons. *Phys. Rev. Lett.* **74**, 738 (1995).
31. Dekorsy, T. *et al.* THz-wave emission by coherent optical phonons. *Physica B Condens. Matter* **219**, 775–777 (1996).
32. Camley, R. E. Long-wavelength surface spin waves on antiferromagnets. *Phys. Rev. Lett.* **45**, 283 (1980).
33. Lüthi, B., Mills, D. L. & Camley, R. E. Surface spin waves in antiferromagnets. *Phys. Rev. B* **28**, 1475 (1983).
34. Damon, R. W. & Eshbach, J. R. Magnetostatic modes of a ferromagnet slab. *Journal of Physics and Chemistry of Solids* **19**, 308–320 (1961).
35. Walker, L. R. Magnetostatic modes in ferromagnetic resonance. *Physical Review* **105**, 390 (1957).
36. Sun, Y. *et al.* Dipolar spin wave packet transport in a van der Waals antiferromagnet. *Nat. Phys.* **20**, 794–800 (2024).
37. Feurer, T., Vaughan, J. C. & Nelson, K. A. Spatiotemporal coherent control of lattice vibrational waves. *Science (1979).* **299**, 374–377 (2003).
38. Cavalleri, A. *et al.* Tracking the motion of charges in a terahertz light field by femtosecond X-ray diffraction. *Nature* **442**, 664–666 (2006).
39. Van Kranendonk, J. & Van Vleck, J. H. Spin waves. *Rev. Mod. Phys.* **30**, 1 (1958).
40. Prabhakar, A. & Stancil, D. D. *Spin Waves: Theory and Applications*. vol. 5 (Springer, 2009).
41. Awschalom, D. D. *et al.* Quantum engineering with hybrid magnonic systems and materials. *IEEE Transactions on Quantum Engineering* **2**, 1–36 (2021).
42. Savitzky, B. H. *et al.* Image registration of low signal-to-noise cryo-STEM data. *Ultramicroscopy* **191**, 56–65 (2018).
43. Handa, T. *et al.* Spontaneous exciton dissociation in transition metal dichalcogenide monolayers. *Sci. Adv.* **10**, eadj4060 (2024).

44. Tulyagankhodjaev, J. A. *et al.* Room-temperature wavelike exciton transport in a van der Waals superatomic semiconductor. *Science (1979).* **382**, 438–442 (2023).
45. Baxter, J. M. *et al.* Coexistence of Incoherent and Ultrafast Coherent Exciton Transport in a Two-Dimensional Superatomic Semiconductor. *J. Phys. Chem. Lett.* **14**, 10249–10256 (2023).
46. Tang, P. & Bauer, G. E. W. Electric analog of magnons in order-disorder ferroelectrics. *Phys. Rev. B* **109**, L060301 (2024).
47. Zhuang, S. & Hu, J.-M. Role of polarization-photon coupling in ultrafast terahertz excitation of ferroelectrics. *Phys. Rev. B* **106**, L140302 (2022).
48. Jackson, J. D. *Classical Electrodynamics*. (John Wiley & Sons, 2021).
49. Scrymgeour, D. A., Gopalan, V., Itagi, A., Saxena, A. & Swart, P. J. Phenomenological theory of a single domain wall in uniaxial trigonal ferroelectrics: Lithium niobate and lithium tantalate. *Physical Review B—Condensed Matter and Materials Physics* **71**, 184110 (2005).
50. Phillpot, S. R. & Gopalan, V. Coupled displacive and order–disorder dynamics in LiNbO 3 by molecular-dynamics simulation. *Appl. Phys. Lett.* **84**, 1916–1918 (2004).
51. Passler, N. C. & Paarmann, A. Generalized 4× 4 matrix formalism for light propagation in anisotropic stratified media: study of surface phonon polaritons in polar dielectric heterostructures. *Journal of the Optical Society of America B* **34**, 2128–2139 (2017).
52. Tayvah, U., Spies, J. A., Neu, J. & Schmuttenmaer, C. A. Nelly: A user-friendly and open-source implementation of tree-based complex refractive index analysis for terahertz spectroscopy. *Anal. Chem.* **93**, 11243–11250 (2021).

## METHODS

### Sample preparation

We exfoliated single crystals of $NbOI_2$, $WO_2Br_2$, and $TaOBr_2$ with the thermal release tape method and transfered flakes on either 1-mm fused silica (THz emission measurements) or a $Si/SiO_2$ substrates. While $NbOI_2$ were transferred under ambient condition owning to its air stability, we exfoliated $WO_2Br_2$ and $TaOBr_2$ in inert environment to minimize chemical degradation. The thicknesses of flakes were determined on either Bruker Dimension FastScan atomic force microscopy (AFM) or an Asylum Research Cypher S AFM. For THz time domain spectroscopy (TDS) measurements, we prepared free-standing samples of $NbOI_2$ and $TaOBr_2$ by the method reported in Ref. [13]. Limited by the size of the parent crystals, we placed $NbOI_2$ and $TaOBr_2$ on a 1-mm pinhole and 500-μm pinhole, respectively, to obtain the free-standing samples. We determine the thickness of the free-standing samples using the method in Ref. [18]. For transmission electron microscopy measurements, $NbOI_2$ crystals were mechanically exfoliated onto poly(dimethylsiloxane) in a nitrogen-filled glovebox. The exfoliated flakes were then transferred from PDMS onto $Si_3N_4$ holey grids (Norcada) using a micromanipulator transfer stage (HW Graphene) under inert conditions and without heating. The grids were cleaned with $O_2$ plasma for 5 minutes immediately prior to transfer to promote flake adhesion.

### Transmission electron microscopy measurements

Selected area electron diffraction (SAED) was performed using a 40 μm aperture, corresponding to an illuminated region approximately 700 nm in diameter on the sample. Measurements were conducted on an FEI Talos F200X microscope operated at 200 kV. Atomic-resolution high-angle annular dark-field scanning transmission electron microscopy (HAADF-STEM) was also carried out on the same instrument at 200 kV, using a convergence semi-angle of 10.5 mrad. To improve image quality, a series of high-frame-rate images (acquired at >1 frame per second) were aligned and averaged, enhancing signal-to-noise while minimizing distortions from sample drift and scan instabilities[42].

### THz emission experiments

THz emission experiments were performed based on our home-built far-field THz setup.[43] We used a Ti:sapphire regenerative amplifier (RA) with a pulse duration of 30 fs, a repetition rate of 10 kHz, and a wavelength centered at 800 nm (Coherent, Legend). The RA output was split into two beams for optical pump and THz sampling. For the determination of THz absorption spectrum, additional broadband THz probe was employed, which was generated by the two-color air plasma method. The pump beam was focused onto sample flakes on a quartz substrate with the pump spot size at the sample being around 50 μm (1/e radius). The pump was incident from the substrate side to characterize the sample response without being affected by the THz absorption of quartz. The polarization of the pump beam was along *b*-axis unless otherwise mentioned. The THz emission was collected/collimated using a parabolic mirror and passed through a high-resistivity Si wafer to block the pump, followed by a THz polarizer (PureWave) that sets the detected polarization parallel to the *b*-axis. The time-domain THz fields were then measured using a 1-mm <110> ZnTe crystal with the sampling beam, via electro-optic (EO) sampling. The setup was purged with dry air. The samples were placed in dry air at room temperature, except for cryogenic measurement, which was performed using a cryostat coupled to a closed cycle Helium recirculating system (Janis RGC4). Prior to the THz experiments, the crystal orientation of flakes was determined using visible polarimetry equipped inside the THz setup.[18]

**Transient reflectance measurements and analysis**

The transient reflectance experiments utilize femtosecond laser pulses (400 kHz, 1030 nm, 200 fs) generated by a solid-state laser (Light Conversion Carbide). The laser output is split into two beams to form the pump and probe arms. The pump beam is routed through noncollinear optical parametric amplifier to generate 700 nm and then to a motorized delay stage to adjust the time delay, Δt. After passing through a filter, the pump beam is modulated by an optical chopper to create alternating pump-on and pump-off signals. For the probe, part of the fundamental beam is focused into a YAG crystal to generate a broadband continuum, which is spectrally filtered to 550 – 680 nm. The probe beam is expanded by a lens for widefield illumination. The pump and probe beams are combined collinearly and directed onto the sample through a 100X, 0.75 NA objective. The pump and probe spot sizes are approximately 1.6 μm and 28 μm, respectively. The same objective is used to collect the reflected light, which is spatially filtered by confocal imaging system equipped with a dual-axis mirror scanning system and the effective probe spot size is 1.1 μm. A short pass filter removes the pump component before the signal is dispersed onto a visible CCD camera (PyLoN-400, Princeton Instruments). Transient reflectance signals ($\Delta R/R_0$) are calculated from the pump-on and pump-off spectra at varying Δt, where $\Delta R = R_{on} - R_{off}$, with the subscript denoting the pump state. To enhance the sensitivity to coherent oscillations, $\Delta R/R_0$ is averaged over all probe wavelengths. All measured were carried out in a closed cycle liquid helium microscope cryostat (Montana Instruments).

To extract the coherence time, we use time domain fits for propagating $|\Delta x| > 0$ with a model $\frac{\Delta R(t)}{R_0} = A\left(\frac{1+erf\left(\frac{t-t_0}{t_r}\right)}{2}\right) sin\, 2\pi f(t-t_0)\, e^{-(t-t_0)/\tau_{coh}}$ , where $\tau_{coh}$ is coherence time, $A$ is amplitude, $t_0$ is wave packet arrival time, $t_r$ is rise time. The group velocity is frequency dependent so that different arrival time within the momentum range excited may contribute to the extended amplitude decay. We use FT intensity in the analysis of transient reflectance in Fig. 2c and 2d for precision in the peak frequencies.

**stroboSCAT measurements and analysis**

A similar setup as that in Refs. [44,45] was used for stroboSCAT measurements. The sample (240 µm thick) was mounted in a closed-loop optical cryostat (Montana Instruments: Cryostation 100) with an in-vacuum objective (Zeiss: 0.9 NA, 100×) and the stage temperature was set to 4 K. The pump pulse (515 nm) was obtained from the second harmonic of the fundamental of a 40 W Yb:KGW ultrafast laser system (Light Conversion Carbide: 40 W @ 1 MHz) and was modulated by an optical chopper at 637 Hz. Linear polarization was ensured by a thin film polarizer (Thorlabs: pump = LPVISC100-MP2, probe = LPVISE100-A) and controlled manually by achromatic half-wave plates (Thorlabs: AHWP10M-600). The probe was obtained from short-passed white light (< 600 nm; Thorlabs: FES0600), generated by focusing ~0.8 µJ of the 1030 nm fundamental into a 5 mm thick YAG crystal (EKSMA Optics). Pump and probe polarizations were aligned along the polar *b*-axis of $NbOI_2$.

The pump pulse was focused on the sample plane, whereas the probe was focused into the back focal plane of the objective to provide widefield illumination (Fig. 3a). The backscattered and reflected probe light was collected through the same objective and was separated from the excitation path through a 50:50 beamsplitter near the entrance pupil of the objective. In the imaging path (after sample interaction), the probe was filtered by a 532 ± 4 nm bandpass filter (Thorlabs: FLH532-4) and imaged on a CMOS array camera (FLIR: BFS-U3-28S5M-C). Differential pump on/pump off frames lead to stroboSCAT images, which were analyzed using code written Python and the SciPy library.

To analyze raw stroboSCAT movies, following subtraction of the initial (-0.50 ps) and final (+50.00 ps) frames, we applied a short-time Fourier transform (STFT — Window size = ¼ data length; Overlap = Window size-1; Zero Padding = 16 data length) to every pixel within the displayed field of view. A Hanning window centered at 3.1325 ± 0.0025 THz was used to bandpass the STFT prior to inversion for Fig. 3b, with images at fine delay time steps (50 fs) around Δt = 20 ps in Fig. 3c. The center of the Hanning window was varied between 3.0075 and 3.2175 THz and select frequency bands are shown at Δt = 40 ps in Extended Data Fig. 9a,b. The amplitude of the filtered STFT (at 3.1325 ± 0.0025 THz) is displayed in Fig. 3d. For every pixel, a 4x4 binned selection was taken from the STFT amplitude movie, and the time delay at which the intensity peaks was extracted. The linear fit of pixel position vs time along a rectangular selection along the propagation direction is used to obtain the velocity of the propagating ferrons (Fig. 3e).

**Derivation of ferron dispersions**

To understand anisotropic transport and thickness dependent group velocity, we investigate the model proposed by Bauer and coworkers[4–7,25,26,46]. Note the equations in this section are labeled as Extended Data (ED) #. We consider the free energy of ferroelectric order below the critical temperature,

$$F = \frac{\alpha}{2} P_x^2 + \frac{\beta}{4} P_x^4 - \boldsymbol{E} \cdot \boldsymbol{P} \qquad (ED.1)$$

where ferroelectric order is polarized along the x-axis, Landau parameters are $\alpha < 0$ and $\beta > 0$, and $\boldsymbol{E}$ is the electric field. At equilibrium condition with zero external electric field, $\frac{\partial F}{\partial \boldsymbol{P}} = 0$, we obtain $P_{x0} = \sqrt{-\alpha/\beta}$; $P_{y0} = P_{z0} = 0$.

Analogous to Landau-Lifshitz equations for the magnon case, Landau-Khalatnikov-Tani (LKT) equations for the ferron case[5,7,28,47] can be written as

$$\frac{1}{\varepsilon_0 \omega_p^2} \frac{\partial^2 \boldsymbol{P}}{\partial t^2} + \gamma \frac{\partial \boldsymbol{P}}{\partial t} = \boldsymbol{E}_{eff} \qquad (ED.2)$$

where $\gamma$ is the phenomenological damping constant, $\boldsymbol{E}_{eff}$ is the sum of effective fields from spontaneous polarization and dynamical electric field, $\boldsymbol{E}_{eff} = -\frac{\partial F}{\partial \boldsymbol{P}} + \boldsymbol{E}(t)$, $\varepsilon_0$ the vacuum permittivity, and $\omega_p$ the ionic motion frequency.

For the harmonic approximation of polarization fluctuations to the first order ($\boldsymbol{p} \ll \boldsymbol{P}$), $\boldsymbol{P}(r,t) = \left(P_0 + \Delta P_x e^{-i\omega t}\right)\hat{x} + \left(\Delta P_y e^{-i\omega t}\right)\hat{y} + \left(\Delta P_z e^{-i\omega t}\right)\hat{z}$, and the dynamical electric field $\boldsymbol{E}(r,t) = (E_x e^{-i\omega t})\hat{x} + (E_y e^{-i\omega t})\hat{y} + (E_z e^{-i\omega t})\hat{z}$, and with the condition $P_0 = \sqrt{-\alpha/\beta}$, we have, $\boldsymbol{E}_{eff} = (-\alpha\Delta P_x - 3\beta P_0^2 \Delta P_x + E_x)e^{-i\omega t}\hat{x} + (-\lambda_1 \Delta P_y + E_y)e^{-i\omega t}\hat{y} + (-\lambda_2 \Delta P_z + E_z)e^{-i\omega t}\hat{z}$. Thus, Eq. (ED.2) is reduced to

$$\left(-\frac{\omega^2}{\omega_p^2} - 2\varepsilon_0\alpha - i\varepsilon_0\gamma\omega\right)\Delta P_x = \varepsilon_0 E_x \qquad (ED.3)$$

$$\left(-\frac{\omega^2}{\omega_p^2} + \varepsilon_0\lambda_1 - i\varepsilon_0\gamma\omega\right)\Delta P_y = \varepsilon_0 E_y$$

$$\left(-\frac{\omega^2}{\omega_p^2} + \varepsilon_0\lambda_2 - i\varepsilon_0\gamma\omega\right)\Delta P_z = \varepsilon_0 E_z$$

where we define the stiffness of polarization fluctuation as $K_x = -2\varepsilon_0\alpha$, $K_y = \varepsilon_0\lambda_1$, $K_z = \varepsilon_0\lambda_2$. From Eq. (ED.3) and $\boldsymbol{D} = \varepsilon_0\boldsymbol{E} + \boldsymbol{P} = \varepsilon_0\boldsymbol{\varepsilon}\boldsymbol{E}$, we can obtain the dynamical dielectric tensor,

$$\boldsymbol{\varepsilon}(\omega) = \begin{pmatrix} 1 + \dfrac{1}{-\dfrac{\omega^2}{\omega_p^2} + K_x - i\varepsilon_0\gamma\omega} & 0 & 0 \\ 0 & 1 + \dfrac{1}{-\dfrac{\omega^2}{\omega_p^2} + K_y - i\varepsilon_0\gamma\omega} & 0 \\ 0 & 0 & 1 + \dfrac{1}{-\dfrac{\omega^2}{\omega_p^2} + K_z - i\varepsilon_0\gamma\omega} \end{pmatrix} \qquad (ED.4)$$

Within the electrostatic limit, $\nabla \times \boldsymbol{E} = 0$ and we define electric potential $\psi$ as $\boldsymbol{E} = -\nabla\psi$. With $\nabla \cdot \boldsymbol{D} = 0$, we have

$$\psi(\boldsymbol{r},t) = \frac{-1}{4\pi\varepsilon_0}\int d\boldsymbol{r}' \frac{\nabla' \cdot \boldsymbol{P}(\boldsymbol{r}',t)}{|\boldsymbol{r} - \boldsymbol{r}'|} \qquad (ED.5)$$

where electric potential is long-range integration of the polarization distribution, corresponding to long-range dipolar interactions[48]. Using the relation in Eq. (ED.3), we obtain the equation as follows

$$\varepsilon_{xx}\frac{\partial^2\psi}{\partial x^2}+\varepsilon_{yy}\frac{\partial^2\psi}{\partial y^2}+\varepsilon_{zz}\frac{\partial^2\psi}{\partial z^2}=0 \qquad (ED.6)$$

For propagating waves within the sample, $\psi = ae^{i\boldsymbol{q}\cdot\boldsymbol{r}}$, Eq. (ED.5) is reduced to

$$\varepsilon_{xx}{q_x}^2+\varepsilon_{yy}{q_y}^2+\varepsilon_{zz}{q_z}^2=0 \qquad (ED.7)$$

For simplicity, we assume $K_y = K_z = 0$, since spontaneous polarization at equilibrium is mainly determined by $\alpha$ and $\beta$. By solving Eq. (ED.7) and for $\varepsilon_0\gamma\omega \ll 1$, we obtain the dispersion relation

$$\omega(\boldsymbol{q})=\frac{\omega_p}{\sqrt{2}}\sqrt{(1+K_x)\pm\sqrt{(1-K_x)^2+\frac{4K_xq_x^2}{q_x^2+q_y^2+q_z^2}}} \qquad (ED.8)$$

We focus on the upper branch of the dispersion, that is, plus sign, which is the forward-moving wave along the polar $x$-axis. For two extrema, i) the Brillouin zone center, $q_x = q_y = 0$ gives $\omega = \omega_p$, ii) large $q_x$ condition ($\gg q_y, q_z$) gives $\omega = \omega_p\sqrt{1+K}$.

Since we focus on thin material slabs, confinement effects on the ferron mode should be considered. For the region outside ferroelectric layer thickness $d$, Maxwell's equations with electrostatic potential give

$$\frac{\partial^2\psi}{\partial x^2}+\frac{\partial^2\psi}{\partial y^2}+\frac{\partial^2\psi}{\partial z^2}=0 \qquad (ED.9)$$

We redefine the electric potential inside and outside the sample

$$\psi^i(\boldsymbol{r},t)=(A\sin q_zz+B\cos q_zz)e^{i(q_xx+q_yy)},\quad |z|\le\frac{d}{2} \qquad (ED.10)$$

$$\psi^o(\boldsymbol{r},t)=\begin{cases}Ce^{-\kappa_zz}e^{i(q_xx+q_yy)}, & z>\frac{d}{2}\\ De^{\kappa_zz}e^{i(q_xx+q_yy)}, & z<-\frac{d}{2}\end{cases} \qquad (ED.11)$$

From Eq. (ED.9) and (ED.11), we have $\kappa_z^2 = q_x^2 + q_y^2$. Eq. (ED.6), (ED.10), and (ED.11) should satisfy boundary conditions that wavefunctions and normal components of $\boldsymbol{D}$ are continuous at $z = \pm d/2$.

From $\psi^i = \psi^o$, we have

$$A\sin\frac{q_z d}{2} + B\cos\frac{q_z d}{2} = Ce^{-\sqrt{q_x^2+q_y^2}d/2} \qquad (ED.12)$$

$$-A\sin\frac{q_z d}{2} + B\cos\frac{q_z d}{2} = De^{-\sqrt{q_x^2+q_y^2}d/2} \qquad (ED.13)$$

From $D_z^i = D_z^o$, we have

$$\left(1-\frac{\omega_p^2}{\omega^2}\right)\left(Aq_z\cos\frac{q_z d}{2} - Bq_z\sin\frac{q_z d}{2}\right) = -\sqrt{q_x^2+q_y^2}Ce^{-\sqrt{q_x^2+q_y^2}d/2} \qquad (ED.14)$$

$$\left(1-\frac{\omega_p^2}{\omega^2}\right)\left(Aq_z\cos\frac{q_z d}{2} + Bq_z\sin\frac{q_z d}{2}\right) = \sqrt{q_x^2+q_y^2}De^{-\sqrt{q_x^2+q_y^2}d/2} \qquad (ED.15)$$

Solving Eq. (ED.12), (ED.13), (ED.14) and (ED.15), we have

$$\left(q_z^2\left(1-\frac{\omega_p^2}{\omega^2}\right)^2 - q_x^2 - q_y^2\right)\tan q_z d = 2\sqrt{q_x^2+q_y^2}q_z\left(1-\frac{\omega_p^2}{\omega^2}\right) \qquad (ED.16)$$

While Eq. (ED.16) is not analytical, numerically solving Eq. (ED.8) and (ED.16) will yield manifolds of in-plane ferron dispersion. In other words, the finite sample thickness $d$, quantizes the wavevector along the out-of-plane direction, $q_z$, which gives rise to multiple bands of ferron modes from Eq. (ED.8).

Since we experimentally measure the wavepacket that arrive the earliest, we focus on the largest group velocity $v_g$ (the most dispersive region of ferron dispersion), which corresponds to: i) polar axis direction, $q_x$ ($q_y = 0$); ii) the condition where $q_x$ and $q_z$ are similar orders of magnitude. For $K = 0.063$ much less than unity, the upper branch of Eq. (ED.8) can be approximated as

$$\omega(\boldsymbol{k}) = \omega_p\left(1+\frac{K_x}{2}\frac{q_x^2}{q_x^2+q_z^2}\right) \qquad (ED.17)$$

The largest $v_g$ can be obtained at $q_x = |q_z|/\sqrt{3}$, $\omega \sim \omega_p\left(1+\frac{K_x}{8}\right)$

$$v_g \sim \frac{\sqrt{3}\omega_p K_x}{8|q_z|} \qquad (ED.18)$$

which shows $v_g \propto 1/q_z$. We approximate Eq. (ED.16) and obtain

$$q_{z,n} = \frac{\kappa_0 + n\pi}{d} \; (n = 0,1,2,3 \ldots) \qquad (ED.19)$$

By comparing Eq. (ED.4) with the dielectric function at room temperature in Fig. 1d, we obtain $\omega_p/2\pi = 3.132$ THz, which corresponds to TO phonon frequency, $K_x$ = 0.063, which gives LO frequency $\omega_p\sqrt{1+K}/2\pi$ = 3.23 THz at large $q_x$ limit. While $K_x$ can be estimated by the Landau parameter $\alpha$, there is no report for $NbOI_2$ in the literature, to our best knowledge. For a widely studied ferroelectric crystal $LiNbO_3$, $\alpha$ is known to be $-2.012 \times 10^9$ $Nm^2/C^2$, which gives $K = -2\varepsilon_0\alpha = 0.036$.[7,8,49] Considering the ferroelectric phase of $NbOI_2$ is robust up to 573 K[13], similar to $LiNbO_3$ [50], we expect $K$ (or $\alpha$) for $NbOI_2$ below room temperature is of the same order as $LiNbO_3$. We estimate $\kappa_0$ = 0.738 ± 0.11 by using Eq. (ED.17) and a fit to $v_g$ of $n = 0$ band in Fig. 2e.

**Transfer matrix method calculations of phonon-polariton dispersions**

We obtain the imaginary part of the reflection coefficient for *p* polarization, Im($r_p$), by the transfer matrix method (TMM)[51]. The dielectric constants along the *b*- and *c*-axes (in-plane axes), Fig. 1b, are obtained from Lorentzian fits to the dielectric constants extracted from the THz data through Nelly[52]. The dielectric constant along the *a*-axis (out-of-plane axis) is assumed to be a constant value of 12. The multilayer structure is vacuum/$NbOI_2$/285nm-$SiO_2$/silicon, where the dielectric constant of the $SiO_2$ and silicon substrate is assumed to be a constant value of 2 and 11.6964, respectively.

The polarization response to an external electric field can also be described macroscopically by the optical dielectric constants of the material. Due to the sharp resonance at the TO phonon frequency along the *b*-axis, the real part of the dielectric constant, Re($\varepsilon_b$), becomes negative between the TO and LO phonon frequencies as shown in Fig. 1b. This leads to the emergence of multiple hyperbolic polariton branches in a thin slab as shown in Fig. 4c. However, to understand the origin of the sharp feature in the dielectric constant spectrum that leads to hyperbolicity, including the oscillator strength and the linewidth of the peak, a microscopic model is required. The agreement between the macroscopic and microscopic approaches suggests that the hyperbolic dispersion features can be well explained by the ferron model with the spontaneous electric polarization in the *b*-axis. Such an agreement does not exist in conventional phonon polaritons because of the absence of spontaneous polarization in the material.

The complex wavevector of the polariton requires solving the poles of the reflection coefficient, $r_p$, in the complex plane, which is challenging, especially for biaxial materials. The maxima of the imaginary part of $r_p$ at real momenta represent a convenient way to determine the polariton dispersion. The imaginary part of $r_p$ implies the degree of phase flip required for the mode to be confined within the sample thickness. In Fig. 4c, Im($r_p$) decreases significantly as frequency approaches that of TO, suggesting that the hyperbolic phonon polariton becomes leaky.

**Data Availability Statement**

The data that support the findings of this study are available in the repository *figshare* at https://doi.org/10.6084/m9.figshare.31895293.

**Methods-only references**

42. Savitzky, B. H. *et al.* Image registration of low signal-to-noise cryo-STEM data. *Ultramicroscopy* **191**, 56–65 (2018).
43. Handa, T. *et al.* Spontaneous exciton dissociation in transition metal dichalcogenide monolayers. *Sci. Adv.* **10**, eadj4060 (2024).
44. Tulyagankhodjaev, J. A. *et al.* Room-temperature wavelike exciton transport in a van der Waals superatomic semiconductor. *Science (1979).* **382**, 438–442 (2023).
45. Baxter, J. M. *et al.* Coexistence of Incoherent and Ultrafast Coherent Exciton Transport in a Two-Dimensional Superatomic Semiconductor. *J. Phys. Chem. Lett.* **14**, 10249–10256 (2023).
46. Tang, P. & Bauer, G. E. W. Electric analog of magnons in order-disorder ferroelectrics. *Phys. Rev. B* **109**, L060301 (2024).
47. Zhuang, S. & Hu, J.-M. Role of polarization-photon coupling in ultrafast terahertz excitation of ferroelectrics. *Phys. Rev. B* **106**, L140302 (2022).
48. Jackson, J. D. *Classical Electrodynamics*. (John Wiley & Sons, 2021).
49. Scrymgeour, D. A., Gopalan, V., Itagi, A., Saxena, A. & Swart, P. J. Phenomenological theory of a single domain wall in uniaxial trigonal ferroelectrics: Lithium niobate and lithium tantalate. *Physical Review B—Condensed Matter and Materials Physics* **71**, 184110 (2005).
50. Phillpot, S. R. & Gopalan, V. Coupled displacive and order–disorder dynamics in LiNbO 3 by molecular-dynamics simulation. *Appl. Phys. Lett.* **84**, 1916–1918 (2004).
51. Passler, N. C. & Paarmann, A. Generalized 4× 4 matrix formalism for light propagation in anisotropic stratified media: study of surface phonon polaritons in polar dielectric heterostructures. *Journal of the Optical Society of America B* **34**, 2128–2139 (2017).

52. Tayvah, U., Spies, J. A., Neu, J. & Schmuttenmaer, C. A. Nelly: A user-friendly and open-source implementation of tree-based complex refractive index analysis for terahertz spectroscopy. *Anal. Chem.* **93**, 11243–11250 (2021).

## Extended Data Figures

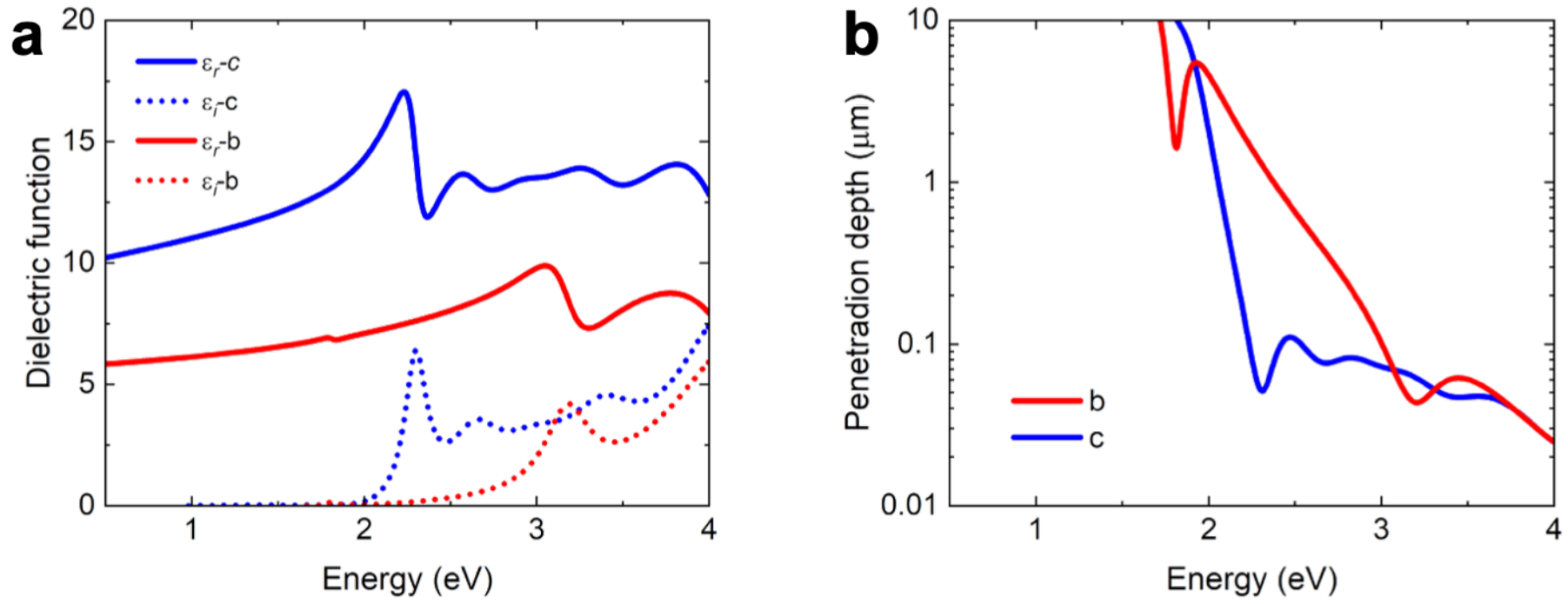

**Extended Data Fig. 1.** Real ($\varepsilon_r$, solid curves) and imaginary ($\varepsilon_i$, dotted curves) dielectric functions and (b) penetration depth experimentally determined from $NbOI_2$ single crystal flakes along the polar b- (red) and non-polar (blue) c-axes at room temperature. The optical dielectric function of $NbOI_2$ was measured using a home-built micro-ellipsometer using a Köhler illumination microscope with a high-numerical aperture (NA = 1.45) objective lens. Angle-resolved spectra are collected by projecting the Fourier (back focal) plane of the microscope on the entrance slit of a spectrometer, enabling highly constrained fitting of dielectric functions on micrometer-sized crystals. The optical measurements are conducted on $NbOI_2$ crystals with known thickness determined by atomic force microscope. The dielectric functions $\varepsilon(\omega)$ are extracted from the experimental transmission and reflection Fresnel coefficients in terms of propagation matrices, and then was globally fitted with the Lorentz oscillators equation:
$\varepsilon(\omega) = \varepsilon_\infty + \sum_\mathrm{i} \frac{A_\mathrm{i}}{\omega_\mathrm{i}^2 - \omega^2 - \mathrm{i}\Gamma_\mathrm{i}\omega}$, where $\varepsilon_\infty$ is the value of the dielectric function at infinite frequency, $\omega_\mathrm{i}$ is the resonant frequency of the oscillator, $A_\mathrm{i}$ is the oscillator strength, and $\Gamma_\mathrm{i}$ is the loss. The accuracy of the dielectric function was verified through transfer matrix modelling of the angle-resolved reflectivity for flakes of varying thickness, showing excellent agreement with experimentally measured angle-resolved reflectance spectra in both thick flakes exhibiting multiple cavity modes and thin flakes.

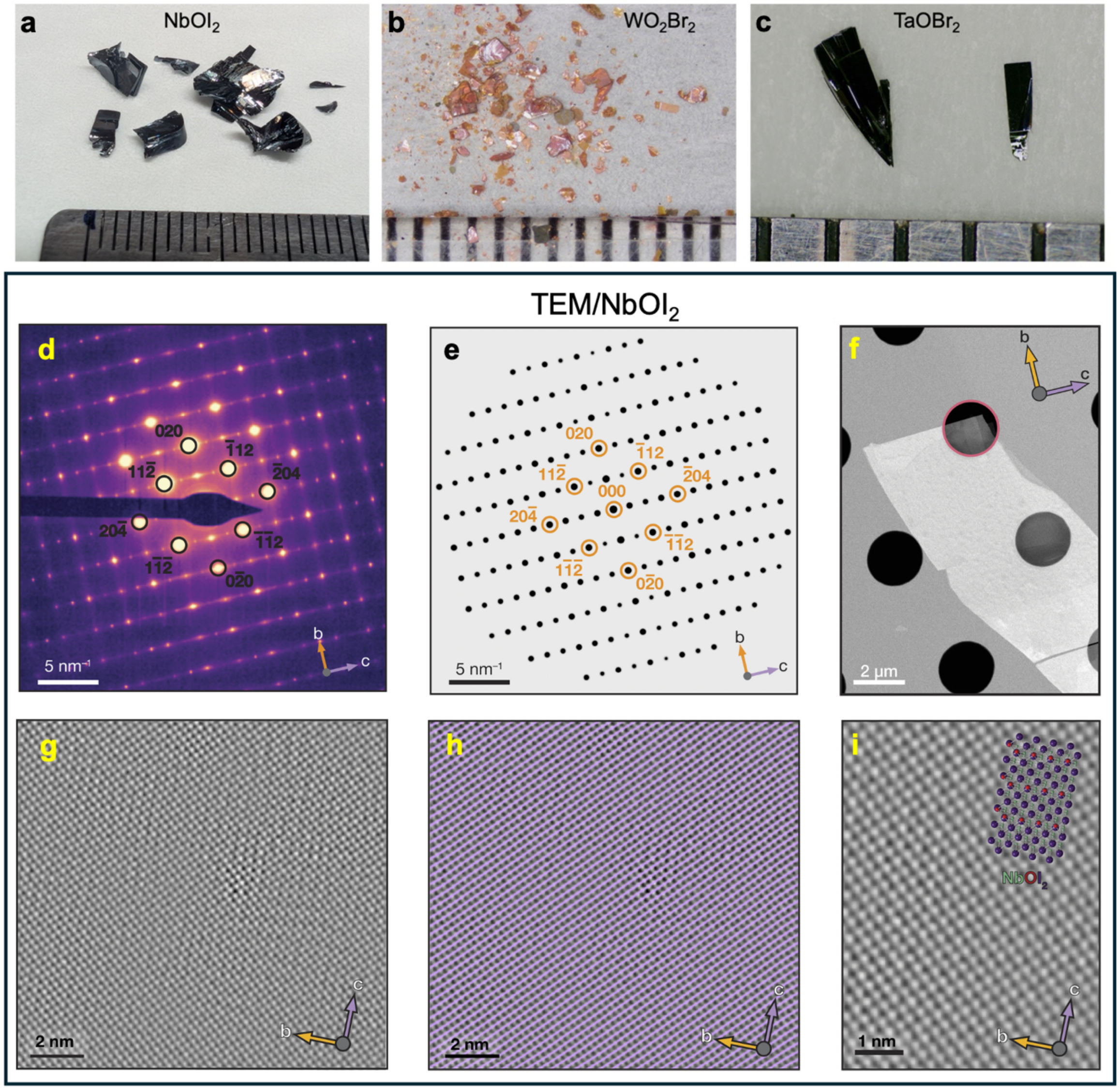

**Extended Data Fig. 2. (a-c)** Optical images of as-grown single crystals of $NbOI_2$ (**a**), $WO_2Br_2$ (**b**), and $TaOBr_2$ (**c**). The scale bar (mm) is shown by the ruler in each photo. **(d-i)** Transmission electron microscopy analysis of $NbOI_2$ flakes. **(d)** Selected area electron diffraction (SAED) pattern of a representative $NbOI_2$ flake taken along the [201] zone axis. **(e)** Simulated diffraction pattern of $NbOI_2$ along the [201] axis. **(f)** High-angle annular dark-field scanning transmission electron microscopy (HAADF-STEM) image of the flake whose diffraction pattern is shown in (**d**). The imaged area is marked with an orange circle. **(g)** Atomic-resolution HAADF-STEM image of the flake shown in (**f**). The image was filtered using a high-pass Bragg filter. **(h)** Image (**g**) overlaid with the inverse Fourier transform generated from the $(\bar{1}12)$ diffraction components of (**g**). The continuous violet $(\bar{1}12)$ fringes are indicative of pristine crystallinity and the absence of grain boundaries. **(i)** Zoomed-in region of (**g**) with the $NbOI_2$ crystal structure overlaid. Crystallographic axes are labeled in all panels, and all data were acquired along the [201] zone axis.

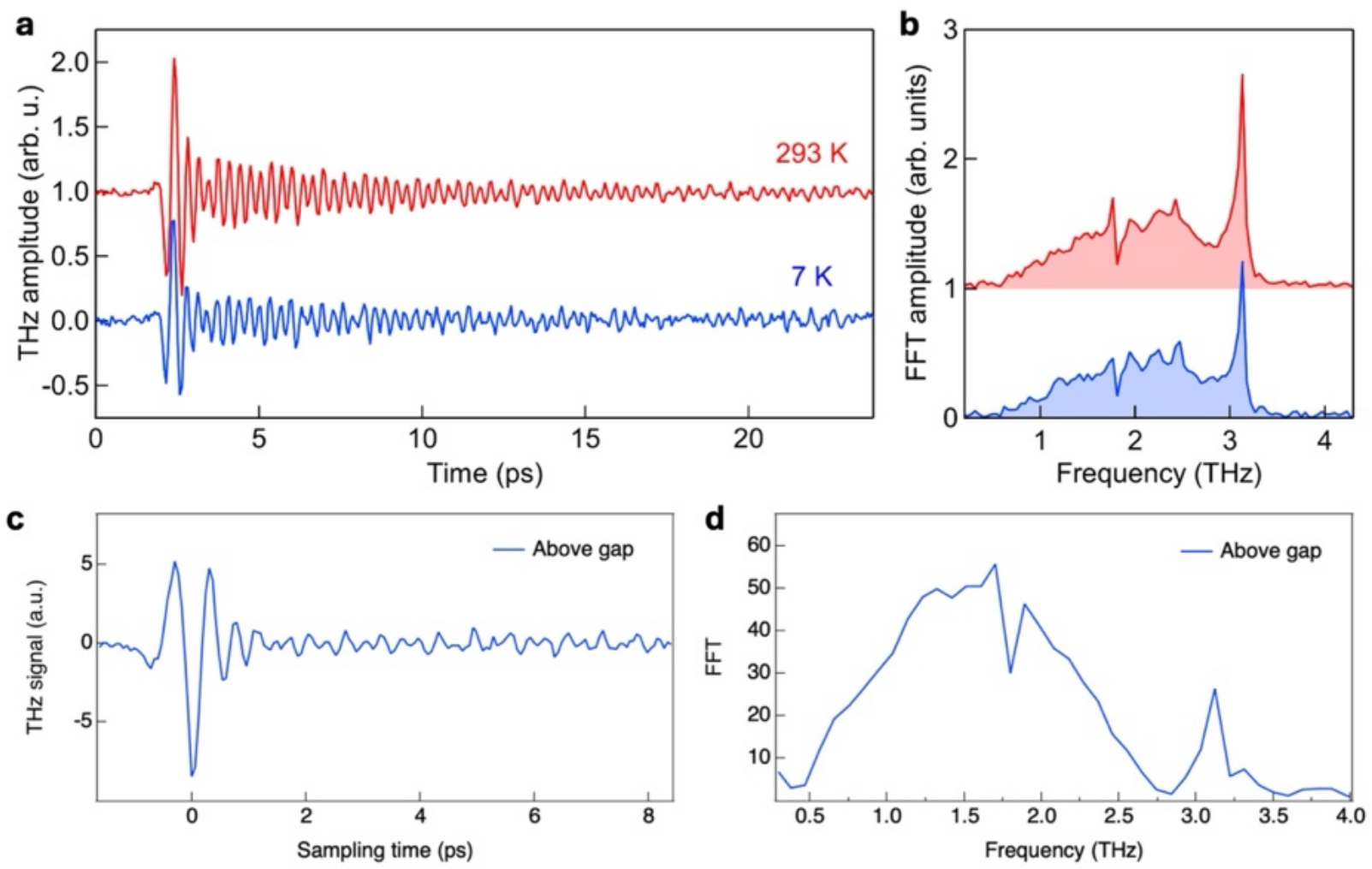

**Extended Data Fig. 3.** (a) Time-domain traces for THz emission from 650 nm thick $NbOI_2$ at the sample temperature of 293 K (red) and 7 K (blue). (b) Fourier transform of data in (a). The pump wavelength and fluence are 800 nm and 1 $mJ/cm^2$, respectively. The 293 K data is vertically offset for clarity. (c, d) THz emission from a 325 nm thick $NbOI_2$ crystal with above gap pump at $h\nu_1$ = 3.1 eV: Time-domain signal (c) and Frequency domain signal (d). The pump wavelength and fluence are 400 nm and 1.2 $mJ/cm^2$, respectively. Compared to below gap pump in (b), the above gap excitation introduces a more intense, broad, and low frequency THz emission (d), which may be assigned to shift-current from charge carrier generation. The 3.13 THz peak at the TO phonon frequency is still clearly observed.

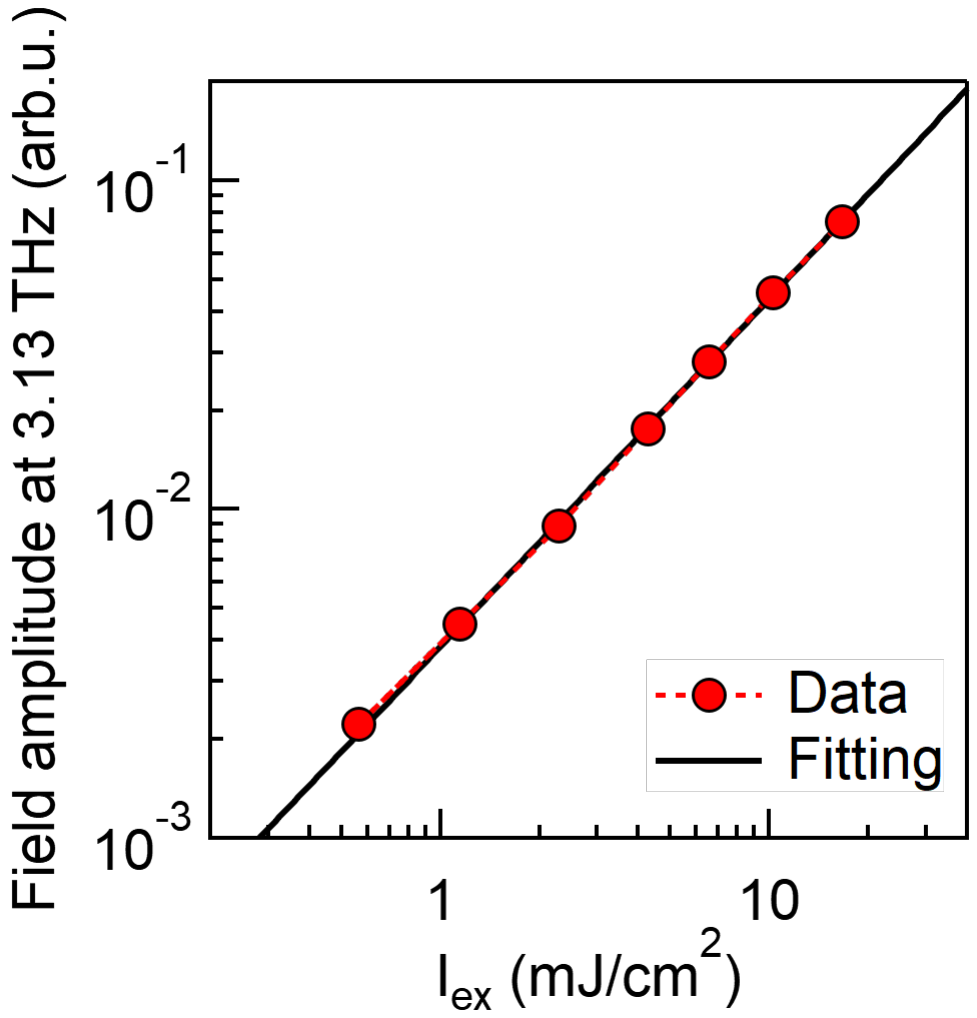

**Extended Data Fig. 4.** Excitation fluence dependence of the emitted field amplitude at 3.13 THz. The black solid line shows fitting to a power law, $(I_{ex})^{\xi}$, giving $\xi = 1.05$, i.e., linear dependence.

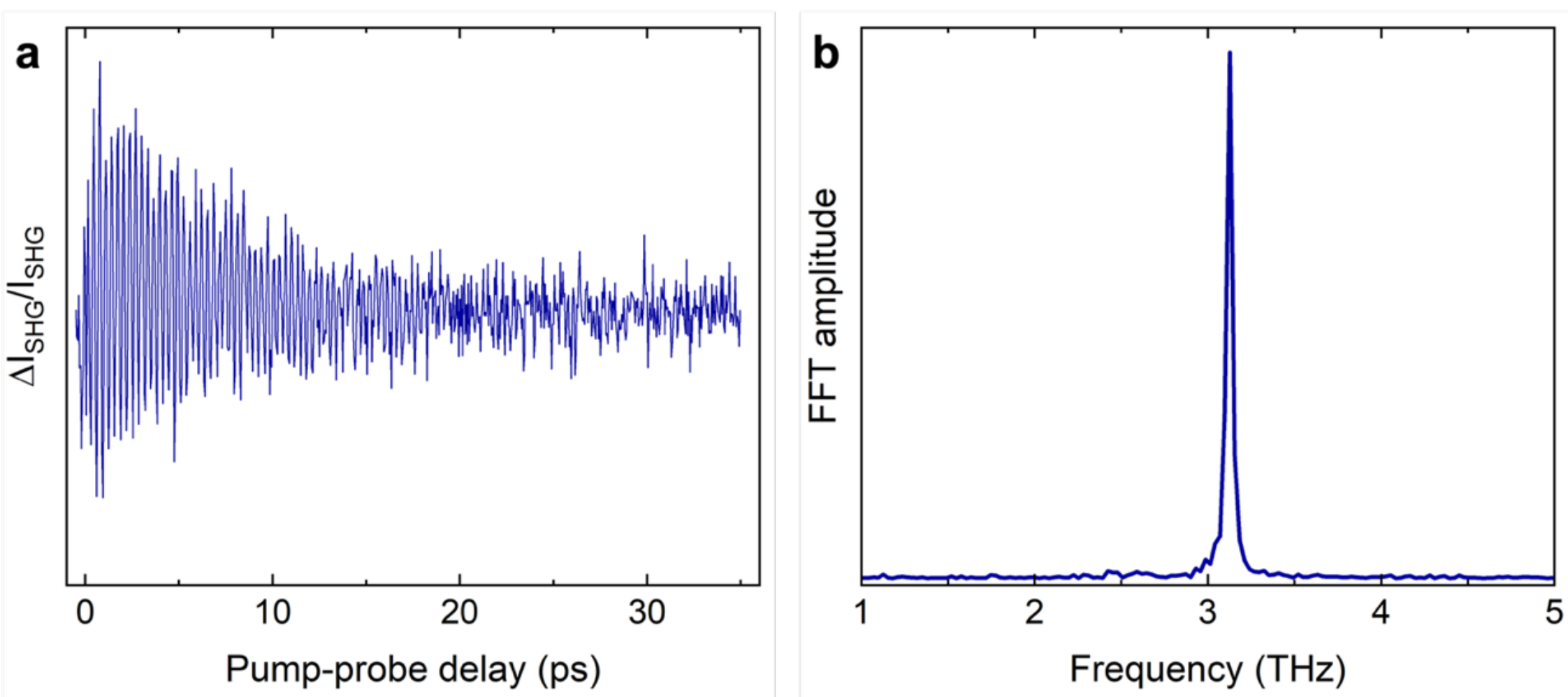

**Extended Data Fig. 5. Time-resolved second harmonic generation.** (a) Time-domain traces for second harmonic generation from 200 nm thick $NbOI_2$ at the sample temperature of 295 K. The pump pulse at $h\nu_1 = 1.75$ eV (fluence, 1.9 mJ/cm$^2$) and the probe pulse at $h\nu_2 = 1.24–1.28$ eV (fluence, 2.7 mJ/cm$^2$) are used. (b) Fourier transform of panel (a).

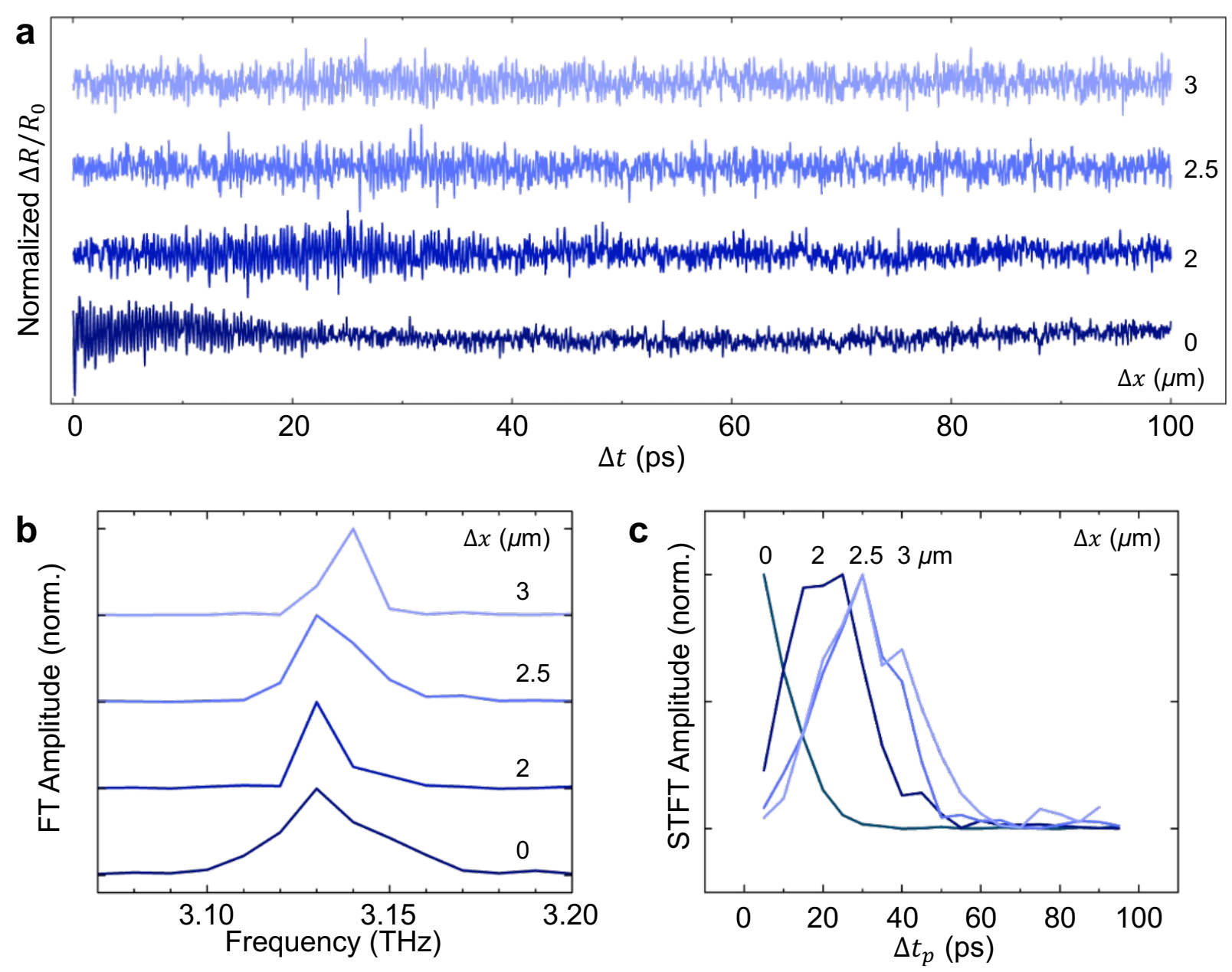

**Extended Data Fig. 6. Coherent ferron propagation at T = 295 K** (a) Transient reflectance $\Delta R/R_0$ (peak intensity normalized) as a function of pump-probe time delay ($\Delta t$) for 200 nm thick $NbOI_2$ at the indicated pump-probe spatial separation along the polar axis, $\Delta x$ = 0, 2, 2.5, 3 μm, respectively. Incoherent background signals are subtracted. (b) Fourier transform of the data for $\Delta x$ = 0, 2, 2.5, 3 μm in panel (a). (c) Short-time window Fourier transform (STFT) of the transient reflectance data in (a) obtained with time window size 10 ps and step 5 ps. The normalized amplitudes of the 3.13 THz signal are shown for $\Delta x$ = 0, 2, 2.5, 3 μm.

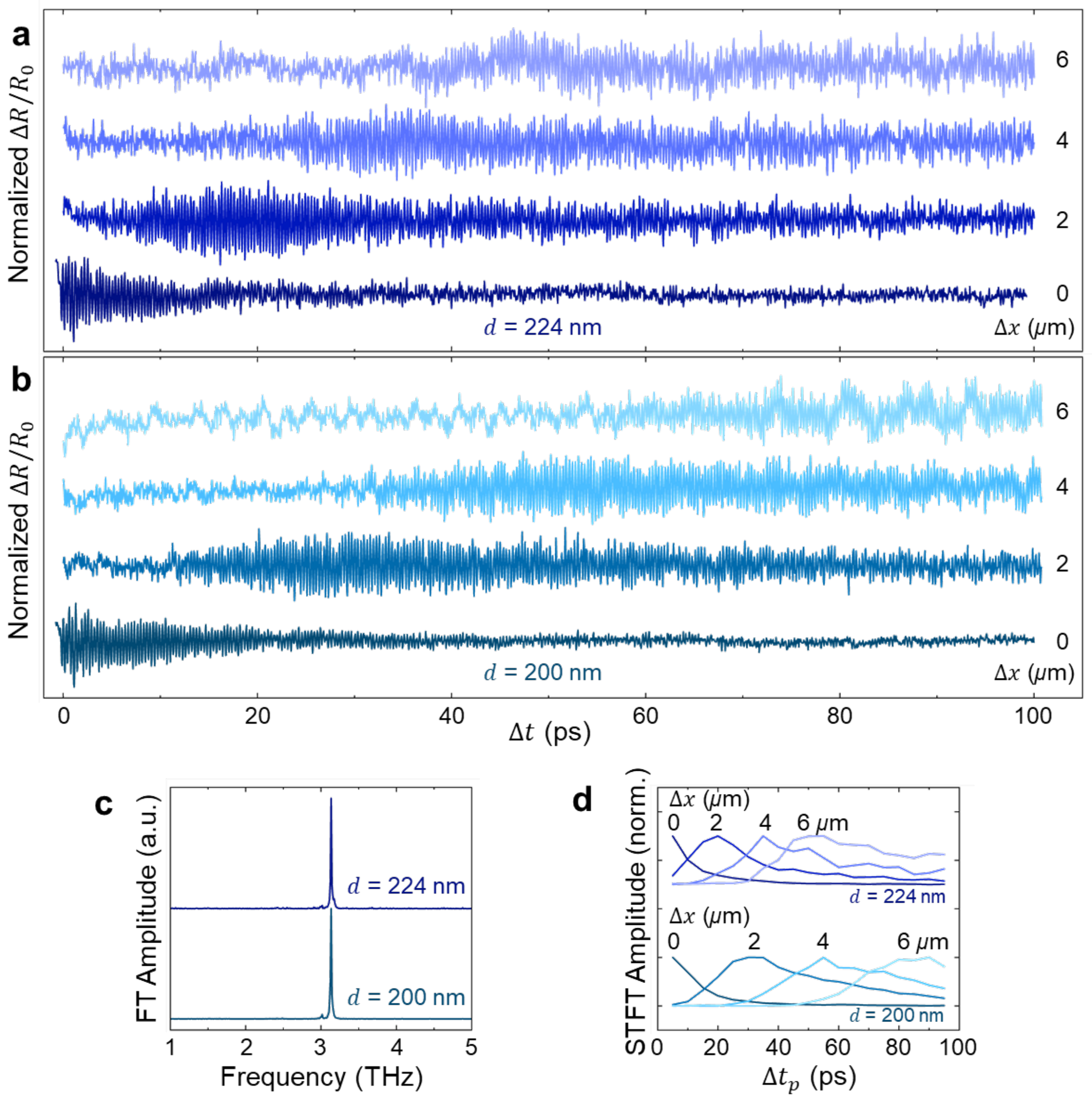

**Extended Data Fig. 7. Polarization wave propagation with above gap pump at $h\nu_1$ = 2.4 eV.** (a,b) Transient reflectance $\Delta R/R_0$ (peak intensity normalized) at 4 K as a function of pump-probe time delay ($\Delta t$) for (a) 224 nm thick and (b) 200nm thick $NbOI_2$ at the indicated pump-probe spatial separation along the polar axis, Δx = 0, 2, 4, 6 μm, respectively. Incoherent background signals are subtracted. (c) Fourier transform of the data for $\Delta x$ = 0, 2, 4, 6 μm in panel (a) and (b). (d) Short-time window Fourier transform (STFT) of the transient reflectance data in (a) and (b) obtained with time window size 10 ps and step 5 ps. The normalized amplitudes of the 3.13 THz signal are shown for $\Delta x$ = 0, 2, 4, 6 μm at the two sample thicknesses, $d$ = 224 nm and 200 nm. All experiments are done at a sample temperature of 3.8 K. The pump pulse at $h\nu_1$ = 2.4 eV (fluence, 0.5 mJ/cm$^2$) and the probe pulse $h\nu_2$ = 1.8−2.22eV (fluence, 32 μJ/cm$^2$) are used.

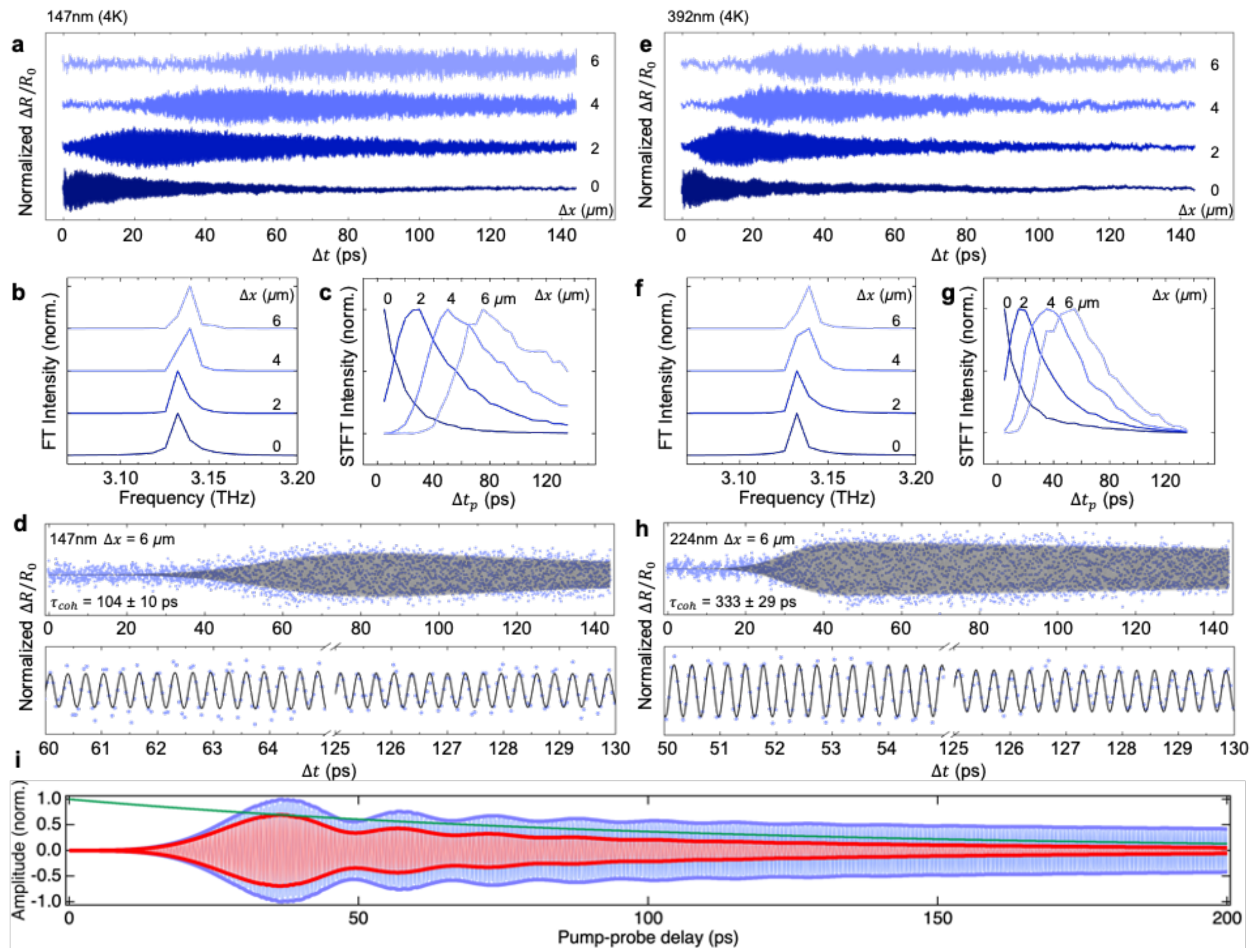

**Extended Data Fig. 8. Polarization wave propagation for 147 nm (a-c) and 392 nm (d-f) thick $NbOI_2$ flakes.** (a), (d) Transient reflectance $\Delta R/R_0$ (peak intensity normalized) at 4 K as a function of pump-probe time delay (Δt) at the indicated pump-probe spatial separation along the polar axis, Δx = 0, 2, 4, 6 μm, respectively. Incoherent background signals are subtracted. (b), (e) Fourier transform of the data for $\Delta x$ = 0, 2, 4, 6 μm in panels (a), (d), respectively. (c), (f) Short-time window Fourier transform (STFT) of the transient reflectance data in (a), (d), respectively, obtained with time window size 10 ps and step 5 ps. The normalized amplitudes of the 3.132 THz signal are shown for Δx = 0, 2, 4, 6 μm. The pump pulse at $h\nu_1$ = 1.77 eV (fluence, 2.8 mJ/cm$^2$) and the probe pulse $h\nu_2$ = 1.82−2.48eV (fluence, 6.5 μJ/cm$^2$) are used. Sample temperature 4K.

We note that in experiment, the upper limit of momentum range excited is $\Delta q$ = 1.73 μm$^{-1}$ for an ideal Gaussian beam at the focal point. Due to deviations from a perfect diffraction-limited pump profile, we expect a smaller momentum range and the corresponding frequency range excited in the $n$ = 0 ferron band less than $\Delta f$ = 43.5 GHz (see Fig. 4b). Other sub-bands ($n$ = 1, 2, …) have narrower excited frequency range than $n$ = 0. We consider that the signal from the pump-probe overlap region ($\Delta x$ = 0) results from the combined contributions from the multiple sub-bands, leading to the observed narrow width. In contrast, the $n$ = 0 band with much higher group velocities is expected to contribute to propagating signal at $\Delta x$ > 0, with finite and distinct momentum windows dominating at different pump-probe separations. Supporting this, we note observable blue-shifts in the ferron frequency with increasing $\Delta x$, Extended Data Figs. 8b,e.

Direct time domain fitting to transient reflectance data at Δt = 6 μm for $NbOI_2$ thickness of 147

nm (g) and 224 nm (h). In (g) or (h), the top panel shows the complete Δt range and the bottom zoomed-in views. The blue dots are data points and black curves are fits to: $\frac{\Delta R(t)}{R_0} = A\left(\frac{1+\mathrm{erf}\left(\frac{t-t_0}{t_r}\right)}{2}\right)\sin 2\pi f(t-t_0)\, e^{-(t-t_0)/\tau}$, where $\tau$ is coherence time, $A$ is amplitude, $t_0$ is wave packet arrival time, $t_r$ is rise time. The fits yield coherence time $t_{coh}$ = 104 ±4 ps and 333 ± 29 ps for thickness 147 nm (g) and 224 nm (h), respectively.

(i) Simulated coherent ferron oscillation at $\Delta x$ = 6 μm for pure dephasing (blue) and dephasing convoluted with population dynamics (red). The green curve is the exponential decay representing population decay.

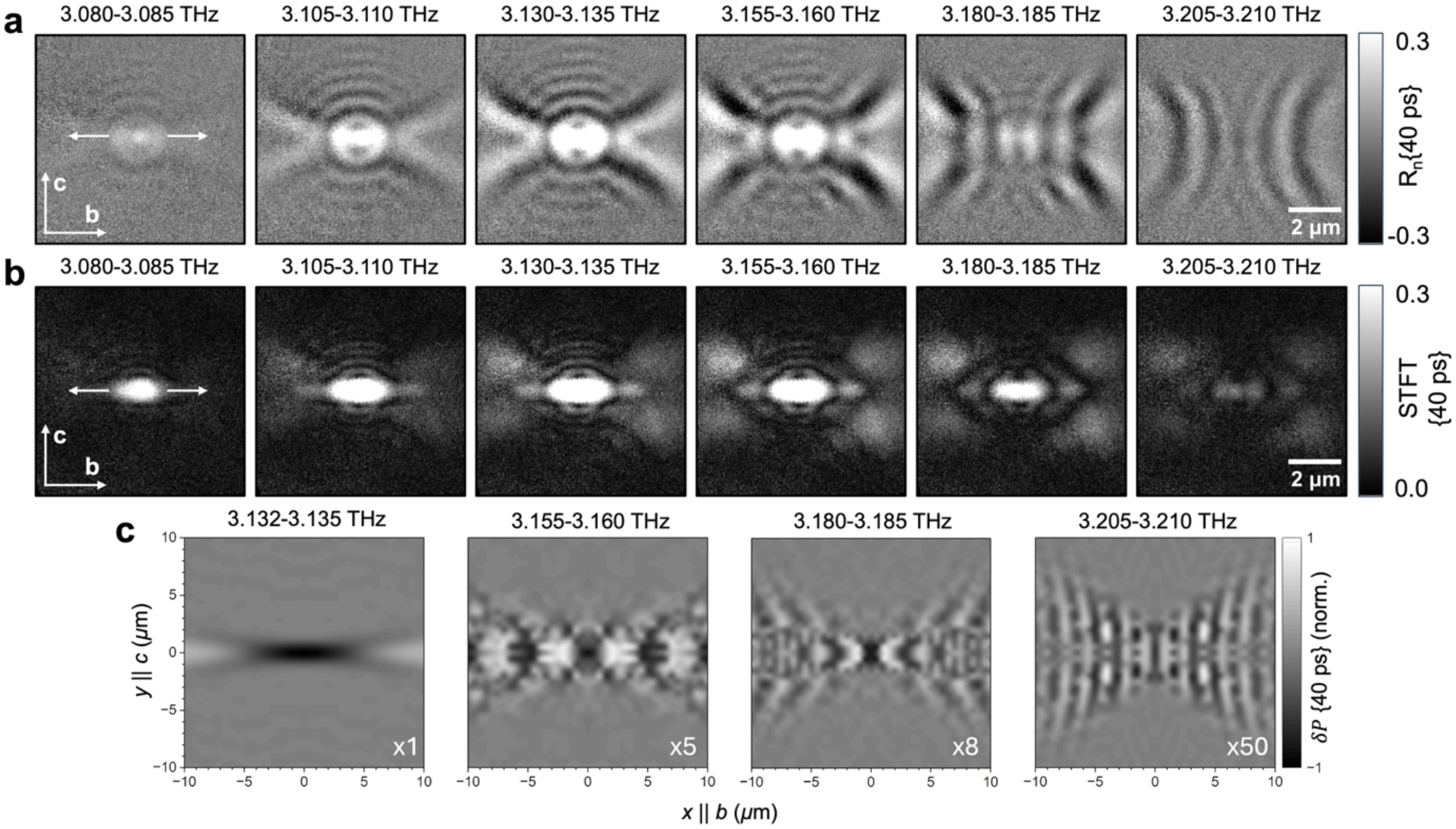

**Extended Data Fig. 9. Frequency dependent ferron maps at time delay 40 ps.** (a) StroboSCAT images for different Fourier bandpass windows. (b) STFT amplitudes from the experiments shown in panel a (normalized against peak signal for 3.130-3.135 THz window). We note that panels a and b include spectral smoothing by 0.08 THz. All scale bars, 2 μm. (c) Calculated spatiotemporal map of ferron transport for different frequency windows.

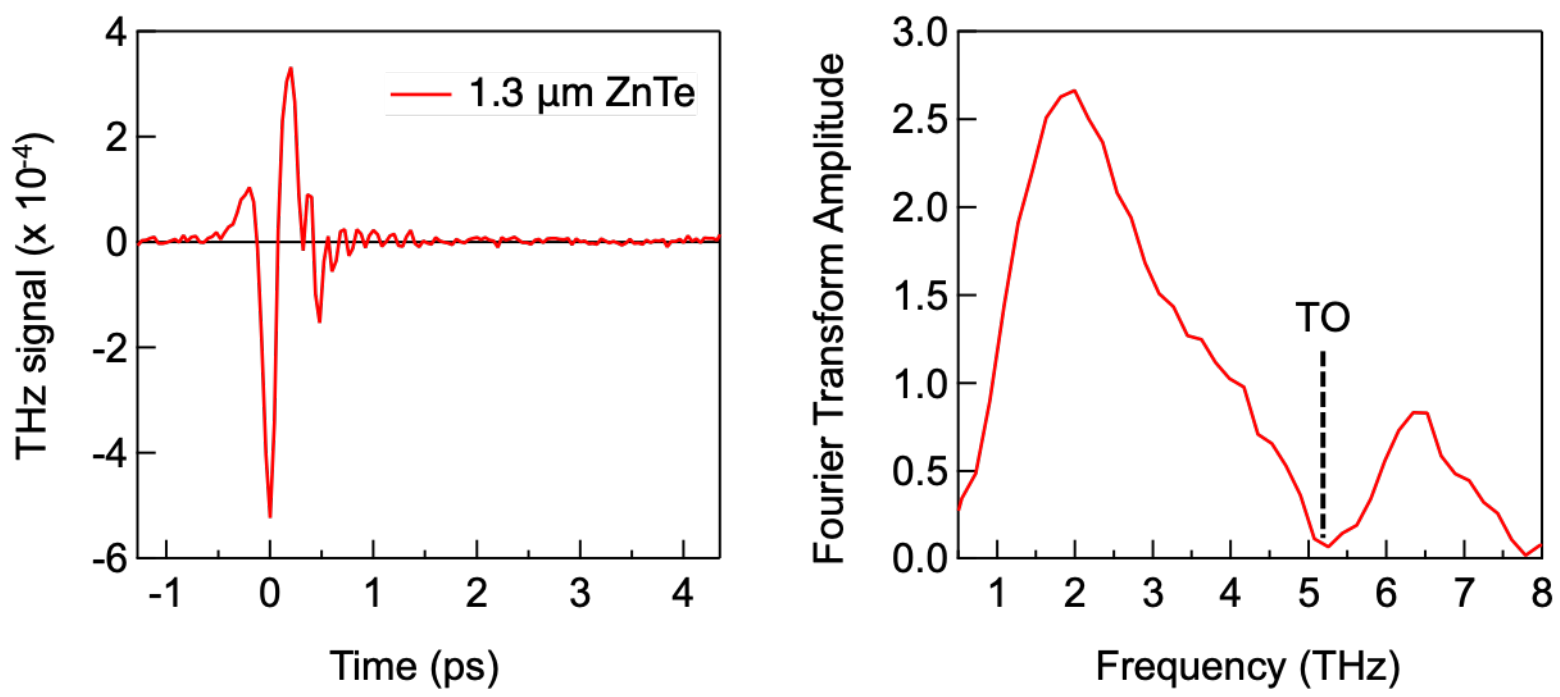

**Extended Data Fig. 10.** THz emission from optical rectification in a thin ZnTe crystal (thickness = 1.3 μm). The detection is via electro-optical sampling in 200 μm thick GaP, and pump laser with the wavelength of 800 nm is used.

**Extended Data Tables**

Table 1. Crystal data and structure refinement for $WO_2Br_2$ at 250 K.

| Empirical formula | $W\ O_2\ Br_2$ |
|---|---|
| Formula weight | 375.67 |
| Temperature | 250 K |
| Wavelength | 0.71073 Å |
| Crystal system | Orthorhombic |
| Space group | *Imm*2 |
| Unit cell dimensions | a = 3.8750(2) Å, α = 90°<br>b = 14.9522(5) Å, β = 90°<br>c = 3.83180(10) Å, γ = 90° |
| Volume | 222.014(15) $Å^3$ |
| Z | 2 |
| Density (calculated) | 5.620 $g/cm^3$ |
| Absorption coefficient | 43.840 $mm^{-1}$ |
| F(000) | 320 |
| Crystal size | 0.11 x 0.03 x 0.004 $mm^3$ |
| θ range for data collection | 2.724 to 36.280° |
| Index ranges | -6<=h<=6, -24<=k<=24, -6<=l<=6 |
| Reflections collected | 3343 |
| Independent reflections | 620 [$R_{int}$ = 0.0449] |
| Completeness to θ = 25.242° | 100% |
| Refinement method | Full-matrix least-squares on $F^2$ |
| Data / restraints / parameters | 620 / 1 / 22 |
| Goodness-of-fit | 1.010 |
| Final R indices [I > 2σ(I)] | $R_{obs}$ = 0.0202, $wR_{obs}$ = 0.0420 |
| R indices [all data] | $R_{all}$ = 0.0207, $wR_{all}$ = 0.0422 |
| Largest diff. peak and hole | 1.039 and -0.930 $e\cdot Å^{-3}$ |
| | |

$R = \Sigma||F_o|-|F_c|| / \Sigma|F_o|$, $wR = \{\Sigma[w(|F_o|^2 - |F_c|^2)^2] / \Sigma[w(|F_o|^4)]\}^{1/2}$ and $w=1/[\sigma^2(Fo^2)+(0.0202P)^2]$ where $P=(Fo^2+2Fc^2)/3$

Table 2. Select bond lengths [Å] for $WO_2Br_2$ at 250 K with estimated standard deviations in parentheses.

| Label | Distances |
|---|---|
| Br(1)-W(1) | 2.4347(7) |
| O(2)-W(1)#2 | 2.1643(6) |
| O(2)-W(1) | 1.7218(7) |
| O(1)-W(1)#1 | 1.757(7) |
| O(1)-W(1)#4 | 2.101(7) |

Symmetry transformations used to generate equivalent atoms:
(1) -x+2,-y+1,z (2) x-1,y,z (3) -x+1,-y+1,z (4) -x+2,-y+1,z-1 (5) x,y,z-1 (6) x+1,y,z (7) x,y,z+1

Table 3. Crystal data and structure refinement for $TaOBr_2$ at 250 K.

| Empirical formula | Ta O $Br_2$ |
|---|---|
| Formula weight | 356.77 |
| Temperature | 250 K |
| Wavelength | 0.71073 Å |
| Crystal system | Orthorhomibic |
| Space group | *Immm* |
| Unit cell dimensions | a = 3.4925(3) Å, α = 90°<br>b = 3.8387(3) Å, β = 90°<br>c = 13.4477(11) Å, γ = 90° |
| Volume | 180.29(3) $Å^3$ |
| Z | 2 |
| Density (calculated) | 6.572 $g/cm^3$ |
| Absorption coefficient | 52.397 $mm^{-1}$ |
| F(000) | 302 |
| Crystal size | 0.132 x 0.039 x 0.002 $mm^3$ |
| θ range for data collection | 3.030 to 29.983° |
| Index ranges | -4<=h<=4, -5<=k<=5, -18<=l<=18 |
| Reflections collected | 2113 |
| Independent reflections | 179 [$R_{int}$ = 0.0721] |
| Completeness to θ = 25.242° | 100% |
| Refinement method | Full-matrix least-squares on $F^2$ |
| Data / restraints / parameters | 179 / 0 / 12 |
| Goodness-of-fit | 1.142 |
| Final R indices [I > 2σ(I)] | $R_{obs}$ = 0.0355, $wR_{obs}$ = 0.0922 |
| R indices [all data] | $R_{all}$ = 0.0375, $wR_{all}$ = 0.0938 |

| Extinction coefficient | . |
|---|---|
| Largest diff. peak and hole | 3.500 and -1.656 e·Å$^{-3}$ |

$R = \Sigma||F_o|-|F_c|| / \Sigma|F_o|$, $wR = \{\Sigma[w(|F_o|^2 - |F_c|^2)^2] / \Sigma[w(|F_o|^4)]\}^{1/2}$ and $w=1/[\sigma^2(Fo^2)+(0.0564P)^2+3.5960P]$ where $P=(Fo^2+2Fc^2)/3$

Table 4. Bond lengths [Å] for $TaOBr_2$ at 250 K with estimated standard deviations in parentheses.

| Label | Distances |
|---|---|
| Ta(1)-Br(1) | 2.668(5) |
| Ta(1)-Br(1)#1 | 2.317(5) |
| Ta(1)-Br(2)#1 | 2.568(6) |
| Ta(1)-Br(2)#2 | 2.888(5) |
| Ta(1)-O(1) | 1.9202(5) |
| Ta(1)-O(1)#2 | 1.9202(5) |

Symmetry transformations used to generate equivalent atoms:
(1) -x,-y,-z (2) -x-1,-y,-z (3) x+1,y,z (4) x,y+1,z (5) -x,-y-1,-z (6) x-1,y,z (7) x,y-1,z

Table 5. Calculated group velocity of acoustic phonons and selected optical phonons of $NbOI_2$ along the crystallographic *b*-axis.

| Phonon mode | Calc. frequency (THz)# | Group velocity (km/s)* |
|---|---|---|
| 1 | 0 | 0.921 |
| 2 | 0 | 2.24 |
| 3 | 0 | 6.44 |
| 7 | 2.381 | $3.24 \times 10^{-3}$ |
| 8 | 2.832 | $71.1 \times 10^{-2}$ |
| 9 | 2.976 | $3.12 \times 10^{-3}$ |
| 10 | 3.129 | $4.82 \times 10^{-2}$ |
| 11 | 3.188 | $1.98\times 10^{-3}$ |
| 12 | 3.415 | $3.13 \times 10^{-4}$ |

#Calculated at the **Γ** point along the **Γ-Y** direction (crystallographic *b*-axis)

*Phonon group velocities of acoustic phonons (mode 1~3) were estimated by calculating the slope, $\frac{\Delta f}{\Delta q}$, where $\Delta f$ represents the frequency difference between the **Γ** point and the nearest $q$ point. Phonon group velocities of selected optical phonons (mode 7-8) were estimated by calculating the slope, $\frac{df}{dq}$, of phonon dispersion curves near the **Γ** point at $q$ = 2 μm$^{-1}$, which corresponds to the largest wavevector our transient reflectance measurements can provide. We fit the phonon dispersion curves near the **Γ** point with a function of $f = a \times q^2 + c$, assuming a quadratic relation between $q$ and $f$ when $q$ is close to the **Γ** point. Data is adapted from Ref. 18.